# Dispersed Matter Planet Project Discoveries of Ablating Planets Orbiting Nearby Bright Stars

Carole A. Haswell[1], Daniel Staab[1,2], John R. Barnes[1], Guillem Anglada-Escudé[3], Luca Fossati[1,4], James S. Jenkins[5], Andrew J. Norton[1], James P.J. Doherty[1], Joseph Cooper[1]

**Some highly irradiated close-in exoplanets orbit stars showing anomalously low stellar chromospheric emission. We attribute this to absorption by circumstellar gas replenished by mass loss from ablating planets. Here we report statistics validating this hypothesis. Among ~3000 nearby, bright, main sequence stars ~40 show depressed 51chromospheric emission indicative of undiscovered mass-losing planets. The Dispersed Matter Planet Project uses high precision, high cadence radial velocity measurements to detect these planets. We summarise results for two planetary systems (DMPP-1 and DMPP-3) and fully present observations revealing a $M_p \sin i = 0.469$ M$_J$ planet in a 5.207 d orbit around the γ-Doradus pulsator HD 11231 (DMPP-2). We have detected short period planets wherever we have made more than 60 RV measurements, demonstrating that we have originated a very efficient method for detecting nearby compact planetary systems. Our shrouded, ablating planetary systems may be a short-lived phase related to the Neptunian desert: i.e. the dearth of intermediate-mass planets at short orbital periods. The circumstellar gas facilitates compositional analysis; allowing empirical exogeology in the cases of sublimating rocky planets. Dispersed Matter Planet Project discoveries will be important for establishing the empirical mass-radius-composition relationship(s) for low mass planets.**

Massive planets in short period orbits were the first exoplanets to be discovered, but their low-mass analogues remained elusive for almost two decades. Recently, however, Kepler results have established there is a significant population of low-mass, ultra-short period (USP) planets: $0.51 \pm 0.07$% of G dwarfs host low-mass USP planets with $P_{orb} < 1$ d[1]. Most of these planets are smaller than 2 R$_\oplus$. Kepler has a narrow field of view, so the planets it discovers are generally distant and difficult to study in detail. Radial velocity (RV) mass determinations of the overwhelming majority of Kepler's small planet discoveries are consequently more-or-less precluded. Nearby analogues of these Kepler USPs must exist, but an enormous effort would be required to find them through targeted RV observations of bright nearby stars with only a 0.5% chance of success for each target. DMPP selects target stars likely to host planets with short orbital periods using archival stellar spectral information. This paper (i) explains the motivation and underlying hypothesis for the DMPP selection criteria, which were developed using inferences from observations of transiting hot Jupiters; (ii) uses the statistics of our first three DMPP planetary system discoveries to test the underlying hypothesis (iii) presents the discovery of DMPP-2 and our RV survey methodology. Companion papers present the discoveries of DMPP-1 and DMPP-3 (*this issue of NA*); the characteristics of these two systems are briefly summarised to place the hypothesis testing in context.

The discovery of an extensive H I exosphere surrounding the hot Jupiter HD 209458 b[2], and subsequent evidence suggests that mass loss is ubiquitous for hot giant planets which orbit close to their host stars. WASP-12 b, a particularly extreme hot Jupiter with $P_{orb} = 1.09$ d[3], is surrounded by an exosphere which overfills the planet's Roche lobe[4,5]. This suggests that the strongly irradiated planet is losing mass[6].

---

[1] School of Physical Sciences, The Open University, Walton Hall, MK7 6AA Milton Keynes, United Kingdom

[2] AVS, Rutherford Appleton Laboratory, Harwell, Oxford, OX11 0QX United Kingdom

[3] School of Physics and Astronomy, Queen Mary University of London, 327 Mile End Rd, E1 4NS London, United Kingdom

[4] Space Research Institute, Austrian Academy of Sciences, Schmiedlstrasse 6, A-8042 Graz, Austria

[5] Departamento de Astronomía, Universidad de Chile, Camino El Observatorio 1515, Las Condes, Santiago, Chile

WASP-12 b's mass loss appears to feed a diffuse circumstellar gas shroud through which we observe the host star[5]. The Mg II h & k line cores of WASP-12 have zero flux. Either the star anomalously completely lacks chromospheric emission, or the emission is absorbed. As a mass-losing planet is present, the latter explanation is more natural and explains the coincidence of two very rare properties in this planetary system. WASP-12's Ca II H & K line core flux is also anomalously low[7], and the interstellar column density along the line of sight is insufficient to depress the stellar chromospheric flux by the observed factor. The hypothesis of absorbing circumstellar gas fed by WASP-12b has been supported by 3D hydrodynamic simulations of the system which show a torus of sufficient optical depth is formed[8].

Stellar activity for FGK stars is commonly parameterised in terms of log $R'_{HK}$, derived from the chromospheric emission in the Ca II H&K line cores relative to the total bolometric emission[9]. Main sequence stars almost invariably exhibit log $R'_{HK}$ > -5.1, a basal level exhibited by inactive solar-type stars, corresponding to the quiet Sun. WASP-12 was an extreme outlier in the distribution of observed log $R'_{HK}$ values[7], with log $R'_{HK}$ = -5.5 and subsequent work revealed that a quarter of stars hosting close-in exoplanets have measured values of log $R'_{HK}$ which lie below the basal limit. These too are apparently viewed through circumstellar gas, originating from planetary ablation, which absorbs the intrinsic stellar chromospheric emission in Ca II H&K[9]. The ongoing OU-SALT survey finds that over 40% of stars hosting transiting planets with semi-major axis $a$ < 0.11 AU lie below the basal limit[10]. For these systems, circumstellar gas shrouds appear to be a common feature; the OU-SALT sample contains stars with log $R'_{HK}$ values as low as WASP-12's.

Intriguingly, about 1.5% of main sequence field stars also lie below the basal limit in log $R'_{HK}$. We postulate these harbour undiscovered close-in ablating planets. To test this hypothesis, we instigated DMPP, studying nearby, bright main sequence stars with log $R'_{HK}$ < -5.1. Figure 1 shows the sample of 2716 main sequence stars from which our targets were drawn (a subset of the then largest available stellar activity compilation[11]; see Methods for details) and the hot Jupiter hosts[12]. The latter are generally more distant and consequently fainter than our sample. This is because transit surveys, which use broad band photometry, sample a larger volume of space than radial velocity surveys which rely on high resolution spectroscopy. The stellar activity compilation we used as our super-sample are generally stars bright enough to be accessible to radial velocity surveys, this is because high resolution spectroscopy or narrow band photometry is required to resolve the Ca II H&K line cores and derive log $R'_{HK}$. Our DMPP targets are the 39 blue points below the basal limit. Figure 2 shows the DMPP targets are drawn from a sample similar in both distance and apparent brightness to RV planet hosts. DMPP targets are ~20 times closer and ~4 mags brighter than the typical known transiting planet host. This is helpful for feasibility of detailed follow-up observations, and is important because interstellar absorption may depress the apparent chromospheric emission of stars at large distances. Interstellar absorption is very unlikely to produce the log $R'_{HK}$ anomaly in any of our targets[13] (See Methods).

Bright, nearby, inactive stars with apparent low activity are preferred targets for RV surveys, so we expected any close-in gas giant planets orbiting the DMPP targets would be already known. Host stars of close-in, small, apparently rocky planets are approximately as rare as main sequence stars with log $R'_{HK}$ < -5.1: Kepler statistics imply ~0.5-1% frequency for the former[1], while we find 1.5% for the latter (see Methods for details). Among the known planet host stars, the anomalously low log $R'_{HK}$ subset includes hosts of low mass planets[9], notably Kepler-68 which hosts a 6 $M_\oplus$ planet on a 5 d orbit. This suggests that the ablation of low mass planets may, like ablation of giant planets, produce absorbing circumstellar gas shrouds. Furthermore, Kepler has revealed spectacular examples of ablation of rocky planets. Kepler-1520b is a catastrophically ablating low-mass rocky planet in a 16 hr orbit around a distant K dwarf[14, 15]. Models to explain the observations of Kepler-1520b invoke a metal-rich vapour co-existing with an optically thick dust cloud ablated from a low mass rocky planet. With $m_V$=16.7, Kepler 1520 is a very challenging target for detailed observational study. DMPP therefore aimed to discover the analogues and progenitors of Kepler 1520b orbiting bright, nearby stars. Such systems are amenable to detailed characterisation.

DMPP performs RV observations sensitive to low mass planets in short period orbits. With an observing run of 7 winter nights, making ~8 HARPS observations per target per night it is possible in principle to detect a ~1 $M_\oplus$ planet in a sub-day orbit around a 0.8$M_\odot$ star. In practice, to resolve period aliases, multiple planets and stellar activity signals, we need a longer baseline. Our firm planet discoveries so far have required multi-season data to establish unambiguous ephemerides. Consequently, most of our DMPP targets are currently only partially investigated. In the companion papers to this article we present two of our first three planetary system discoveries. These are RV discovered planets, but our selection of targets introduces clearly-defined selection effects which are comparably strong to those of transit surveys. Thus, it is advantageous to label our discoveries using the DMPP project acronym so that statistical analyses of exoplanet demographics can pick them out with ease.

# Results

We summarise our findings for DMPP-1 and DMPP-3 (*see this issue of NA for these articles*). We give a full account of the discovery of DMPP-2b: a giant planet orbiting a γ-Doradus pulsator.

## DMPP-1

HD 38677 / DMPP-1 showed a clear reflex RV signal from our first observing run, but it required data spanning 760 d to pin down the periodicities sufficiently to validate it. Even so, some uncertainty remains regarding the correct orbital periods among the aliases. Despite our uncertainty of the true configuration, it is clear that DMPP-1 is a compact multi-planet system containing at least three super-Earth planets in < 10 d orbits, with evidence for a warm Neptune in a 20 d orbit. We discovered the DMPP-1 planetary system with 148 HARPS observations over three observing seasons. This immediately demonstrates the efficiency of our programme; a similar system orbiting HD 215152 required 373 observations taken over 13 years[16]. The factors of ~ 4 in elapsed time, and > 2 in number of observations required are achieved by our bespoke observing strategy, informed by the *a priori* likelihood of short period planets orbiting stars with anomalously low log $R'_{HK}$.

## DMPP-2

We made 49 HARPS RV measurements of DMPP-2 (the F5V star, HD 11231 – see Methods for system parameters) over four observing runs (P95, P97, P98A and P98B) and incorporated 7 unpublished archival measurements using the MIKE spectrograph at the Las Campanas Observatory[17], see Methods for details. Figure 3 shows considerable RV variability within each HARPS run with r.m.s. variability of 30.5 m s$^{-1}$ over the entire HARPS data set. We initially performed a simple search for periodic signals in the HARPS data, allowing no RV offsets between the various observing runs, and assuming eccentricity, $e = 0$. This latter assumption suppresses false positives[18]. Figure 4(b) shows the likelihood periodogram obtained with this RV model. A signal was detected in the P95+P97 data ($P_{orb}$ ~ 5.4 d; FAP = $3.6 \times 10^{-4}$) and adding the P98 data yielded $P_{orb}$ = 5.998 d (FAP of $5.9 \times 10^{-7}$). The window function causes a complex alias structure with many sharp sidelobes. The second peak at $P_{orb}$ = 5.206 d, is only 4.8 times less likely than the main peak, *i.e.*, we cannot reliably distinguish between them[18]. A noise-free Keplerian signal at $P_{orb}$ = 5.998 d or 5.206 d with the sampling of our data reproduces the main alias peaks and sub-peaks[19] of Figure 4b. In particular, the prominent periodogram peaks seen either side of 1 d and 0.5 d are aliases of the $P_{orb}$ ~

6 d signal. All other alias peaks and sidelobes in the periodogram have ΔlogL values indicating they are at least 3000 times less probable than the highest peak.

We noted a large RV dispersion about the maximum likelihood fit indicated by Figure 4(b). The residuals show peak-to-peak variability of up to 85 m s$^{-1}$ on timescales of a few days and the jitter is ~18 m s$^{-1}$.

Systematic RV offsets between runs

In our checks of the line profile behaviour, we discovered a ~350 m s$^{-1}$ change in the FWHM of the cross-correlation function (CCF) as calculated by the HARPS DRS between the P95-P97 and the P98 observing runs. We attribute this to a change in the dominant pulsation modes. Other stars observed during these DMPP campaigns did not show this, ruling out an instrumental origin, so this was a systematic change in the DMPP-2 stellar lines. Since there may be a corresponding systematic RV shift, we treated the P95-P97 and P98 observations as independent runs by allowing separate offsets, $\gamma_{P95+P97}$ and $\gamma_{P98}$, in the maximum likelihood fitting model (see Methods). The resulting periodogram in Figure 4(c) is similar to that in Figure 4(b) but the strongest peak is now $P_{orb}$ = 5.205 d and exceeds the second peak at $P_{orb}$ = 5.998 d by ΔlogL = 4.85 (128 times more probable). Fig. 4 (d) shows the likelihood periodogram with the addition of the MIKE data and allowing for three independent offsets: $\gamma_{P95+P97}$, $\gamma_{P98}$ and $\gamma_{MIKE}$. As illustrated in Fig 1, these data span 12 years. The strongest peak remains $P_{orb}$ = 5.205 d, which now exceeds the second peak by ΔlogL = 6.59 (731 times more probable). The combined HARPS+MIKE data ameliorates effects from our HARPS run lengths which are comparable to the detected period. Our HARPS data alone cannot reliably distinguish between the 5.2 d and 6.0 d periods due to the sampling combined with the jitter from pulsations and the one-time FHWM offset.

Keplerian parameters from posterior samples

The *a posteriori* system parameters (see Methods) derived for each of the three models considered are shown in Table 1 (Columns 1 to 3) with semi-Gaussian eccentricity prior, $\sigma_e$ = 0.05[18]. Column 4 of Table 1 is the solution with an eccentricity prior of $e$ = 0. The *a posteriori* solution with $\gamma_{P95+P97}$ and $\gamma_{P98}$ has $\chi_r^2$ reduced by a factor of 1.49 compared with the case with a single offset, $\gamma_{HARPS}$. The offset between the fitted HARPS $\gamma_{P95+P97}$ and $\gamma_{P98}$ values is 24.3 m s$^{-1}$ (23.1 m s$^{-1}$ when the MIKE data are included). The $P_{orb}$ ~ 5.2 d signal is present in all subsets of data in which we searched for Keplerian signals. Data sampling causes the best fit solution to switch between the ~6.0 d and ~5.2 d sidelobes, depending on the details of the RV model and combination of subsets used. Using all data sets favours the $P_{orb}$ = 5.207 d signal, which is coherent over the >12 yr span of the observations. The fit using all observations with $e$ = 0.078 in Table 1 (column 3) yields the best fit with $\chi_r^2$ = 1.059. The corresponding circular solution with $e$ = 0 (column 4) yields a slightly higher $\chi_r^2$. The phased RV curves for the *a posteriori* solutions using the full data set with and without eccentricity are shown in Figure 5. We find no evidence for further Keplerian signals or RV signals caused by stellar rotation (see Supplementary Information).

γ-Doradus pulsations

The RV jitter in DMPP-2 arises from line profile variability (Supplementary Figs 5 & 6). There is an immediately obvious and self-consistent explanation for the line profile variability: the star lies at the red edge of the space in the HR diagram occupied by γ-Doradus pulsators[21] (Supplementary Table). These stars are A-F type main sequence stars that show multi-periodic photometric and spectroscopic variability[20] attributed to non-radial gravity-mode pulsations[22]. Canonical pulsation periods are between 0.4 d and 3 d, with typical photometric amplitudes of 0.002 - 0.1 mag[23]. Recent work on stars in the Kepler field reveals that the γ-Doradus pulsators lie entirely within the δ-Scuti instability strip[21], suggesting the two types of variable are not distinct. Previously, the δ-Scuti pulsators, which have periods down to 0.5 hours and attributed to p-mode oscillations[23], were thought to occupy a different instability strip in the HR diagram, bluewards of the γ-Doradus region, although with some overlap and hybrid behaviour[24]. It now appears that the two classes arise because different pulsation modes are excited in different stars, rather than being due to fundamentally different driving and damping mechanisms. γ-Doradus pulsations are not particularly rare: estimates indicate they are present in up to ~20% of A7 to F5 type stars[24], though < 2% of stars with the effective temperature of DMPP-2 exhibit γ-Doradus pulsations[21]. The behaviour has, however, been seen in stars as cool as $T_{eff}$ ~ 6300 K, i.e. spectral type F8[25, 26]. The spectral line profile variability of γ-Doradus stars has been studied in detail, giving insights into stellar structure[27-30]. Typical RV amplitudes induced by the pulsations for canonical, well studied examples are 2 - 4 km s$^{-1}$, but Kepler targets identified photometrically as γ Dor pulsators have also been detected showing peak-to-peak RV variability as low as ~100 m s$^{-1}$ [23,31,32]. The DMPP-2 line profile shape parameters exhibit some periodic variability which we attribute to pulsations (Supplementary Fig 4). The periods are distinct from the RV period.

There have been few planet discoveries around pulsating stars because known pulsators are avoided in RV planet search programmes. The transiting hot Jupiter WASP-33 orbits a rapidly rotating A-type star with δ-Scuti pulsations[33,34]. Recently, short-period (~90 min), low-amplitude photometric pulsations were discovered[35] for the eccentric transiting hot Jupiter host HAT-P-2. The pulsations may be caused by tidal star-planet interactions and may cause the observed stellar RV jitter of 36 m s$^{-1}$. Pulsations excited by tidal effects have also been observed in eccentric stellar binary systems[36].

Results from a large HARPS study of 185 A-F type stars, including several pulsators, find that pulsating stars do not show correlations between line cross-correlation profile bisector span (BIS) and RV[37]. Instead, the BIS timeseries is much more variable than the RVs, with median r.m.s. BIS/RV ~3. This leads to a vertical spread of points in the RV-BIS plane. A combination of pulsations and an orbital companion leads both parameters showing large-amplitude variability with no correlation. This is precisely the behaviour in DMPP-2: when considering the residual RVs with the Keplerian signal removed, we find that r.m.s. BIS/RV = 2.8 (Methods; Supplementary Fig 5b). Additional examples are known with stellar rather than planetary companions[38].

The only exoplanet host that has been explicitly identified as a likely γ-Doradus star is the transiting hot Jupiter host WASP-118. The Kepler light curve of this F6 star[31] exhibits low amplitude, multi-periodic pulsations at periods between 1 d and 2.5 d. No pulsation-related analysis of WASP-118's RV and line profile variability has been done to date and the extant data is largely not publicly available[32]. Intriguingly, the BIS-RV plot of WASP-118 does show a composite behaviour similar to DMPP-2's.

There is an RV discovery of a 25 $M_J$ Brown Dwarf orbiting the γ-Doradus A9V star[39], HD 180777. In this case, peak-to-peak RV variations of 1.7 km s$^{-1}$ due to pulsations remain after subtracting the best fit Keplerian signal. The amplitude ratio of the r.m.s. BIS/RV reaches 1.2. The DMPP-2 system could be seen as a scaled-down version of HD 180777, with a lower mass star and companion, combined with lower amplitude pulsations.

Searches for Signatures of Pulsations
We have only 49 HARPS spectra, and γ-Doradus pulsations are generally multiperiodic, but nonetheless we examined the line profiles for signatures. We do not have sufficient data to determine the pulsation periods, but we examined the line profile residuals after combining several thousand absorption lines in each spectrum[40]. A change in line shape from epoch to epoch is evident in the timeseries, while the lines are also clearly variable at a given epoch (Methods; Supplementary Fig 7). Crucially, we find that the line distortions resemble those seen in a γ Doradus star[41].

We find a low amplitude (0.6%) photometric modulation in archival photometry which is consistent with the ~1d period expected for γ Doradus pulsations. It is also consistent, however, with artefacts due to nightly sampling in the single-site SuperWASP data we examined (see Methods for details and

Supplementary Fig 8). Higher quality multi-site or space-based photometry will be required to perform asteroseismology on DMPP-2.

## DMPP-2 Summary

DMPP-2b is a close orbiting giant planet in a $P_{orb}$ = 5.207 d orbit with $M_P \sin i = 0.457^{+0.044}_{-0.034}$ $M_J$. We ascribe the RV residuals to γ-Doradus pulsations, and note that these offer the promise of unusually detailed knowledge of the stellar properties for DMPP-2. The star is on the cool edge of the γ-Doradus region, where only ~2% of stars are known to pulsate. This suggests the possibility that the pulsations are tidally excited by the close-orbiting planet. DMPP-2b is a highly irradiated planet: assuming an albedo of 0.5, the equilibrium temperature is $T_{eq} \approx$ 1000K. It lies near the high-mass, long-period border of the Neptunian desert. It has a particularly luminous host star so it is hotter than most planets at the same location in the ($M_p$, $P_{orb}$) plane. We persisted in RV observations despite the signatures of pulsations in the line profile because we had identified the star *a priori* as a likely host of a close-orbiting planet via absorption attributed to circumstellar gas ablated from the putative planet. DMPP-2b appears to be in the process of significant mass loss; we may be observing it in a short-lived evolutionary phase which sculpts the Neptunian desert. DMPP-2b **and the recently announced discovery of β Pic c are currently the only RV discoveries of planets orbiting strongly pulsating stars.**

Angular momentum considerations suggest that the ablated gas will be concentrated in the orbital plane. Thus DMPP-2b's transit probability exceeds that for randomly oriented orbits, i.e. > 13%. We searched for transits in SuperWASP photometry and can consequently place a preliminary limit on the transit depth δ < 0.6% (see Supplementary Information for details). Targeted photometry will establish whether DMPP-2b transits and will allow the pulsation period(s) to be measured. Since we know there is circumstellar absorption in our line of sight, transmission spectroscopy promises to reveal details of the composition of the planetary mass loss. The combination of this with robustly determined detailed stellar properties derived from asteroseismology mean that DMPP-2 is a particularly promising object for further study, irrespective of whether DMPP-2b transits. For example, DMPP-2 may offer the best prospect for testing the hypothesis that the photospheric abundance patterns of close-in planet hosts are polluted by the accretion of planetary material. In a first step towards this, we performed an abundance analysis using our HARPS data, finding a slight overabundance of some iron peak elements relative to solar (see Methods).

## DMPP-3

Our third discovery has an unexpected system architecture. HD 42936 / DMPP-3 hosts a super-Earth planet in a 6.7 d orbit, along with a companion, DMPP-3B. DMPP-3B has a minimum mass at the boundary between brown dwarfs and low mass stars, and is probably an L dwarf with steady hydrogen burning. It is in an $e$ = 0.59, 507 d orbit. This is a rare configuration. DMPP-3AB is a factor of 2 more compact than any other binary system hosting an S class (circumprimary) planet. It thus presents a challenge to planet formation models: a circumprimary protoplanetary disc would be truncated by DMPP-3B's tidal influence. This limits the mass available to form the super-Earth planet(s) and reduces the lifetime of the disc. Furthermore, planetesimals formed within the circumprimary disc would be vulnerable to scattering by gravitational interactions with DMPP-3B, making them less likely to coalesce to build planets. These considerations explain the dearth of known S class planets in binary star systems with $P_{orb}$ < 2000 d. An alternative scenario has been explored[42], in which S-class planets result from the scattering of a circumbinary planet into an S-type orbit around one of the stars in a binary. This too appears *a priori* unlikely, with only about 1% of scattered circumbinary planets ending up in circumprimary orbits for a system similar to DMPP-3AB[42]. Furthermore, there are no known circumbinary planets for binaries with $0.18 < \frac{a}{a_{DMPP-3}} < 4.3$. It seems that the DMPP-3AB planetary system is the result of an unusual evolutionary pathway. Indeed, it seems likely that the system has undergone Kozai-Lidov migration, and is a super-Earth analogue of hot Jupiters formed in this way. The warm Jupiter system Kepler-693[43] is potentially a similar system in an earlier evolutionary phase, particularly if we hypothesize that DMPP-3Ab was previously more massive and has lost its envelope. While the size of DMPP-3Ab is unknown, its mass suggests it lies below the radius valley[44]. These unusual systems offer powerful opportunities to extend our understanding of the mechanisms which sculpt the Galaxy's planetary system population by examining the outcomes which result from extreme scenarios. Indeed, the DMPP project itself was initially motivated by the properties of the extreme hot Jupiter WASP-12b. DMPP-3B offers an opportunity to calibrate models of L dwarf atmospheres.

# Discussion

## DMPP, Circumstellar Gas, and Exoplanet Demographics

Patterns in the demographics of short period planets[45-48] are due to a significant lack of short period planets at intermediate mass or radius: the Neptunian desert[49]. This feature cannot be due to observational selection effects and has persisted as more exoplanets have been discovered. Possible explanations include the loss of gaseous envelopes via Roche lobe overflow as a result of prodigious radiation from the host star[50], or tidal disruption of planets which arrive in short period orbits via an eccentric orbit and subsequent circularisation[51]. Recent work incorporating planetary evolution models and the demographics of known short orbital period planets concludes both mechanisms combine to create the desert[52].

DMPP targets stars with circumstellar gas shrouds, which is a likely side-effect of photoevaporation or tidal disruption. Figure 6 shows the first six DMPP planets plotted with confirmed planets from the NASA Exoplanet Archive. The Neptunian desert seems to persist up to 12-15 d. Four of the six DMPP planets lie below the desert, while DMPP-2b lies above it. Around the 18.6 d orbital period of DMPP-1b there are too few known planets to sensibly discuss the mass distribution.

None of the first six DMPP planets lie within the Neptunian desert. This is encouraging for the tidal disruption explanation of the desert, which predicts that large (giant) planets and small (rocky) planets arrive at locations below and above the desert after circularisation of eccentric orbits as a consequence of their differing mass-radius relationships. Planets which would have circularised to orbits within the desert are tidally disrupted before the orbit circularises[51]. If the gas shrouds are due to on-going mass loss from a short orbital period planet, we might expect that DMPP planets could be caught in a short-lived phase in which they are moving down in the ($M_P$, $P_{orb}$) plane, transitioning through the Neptunian desert. The radii of planets below the desert implies they are not bare solid rocky bodies, they appear to retain bound envelopes[53]. DMPP cannot yet provide sufficient statistics to determine the mechanism causing the circumstellar gas in these systems, or determine the mechanism(s) creating the Neptunian desert. Full exploration of the remaining DMPP targets is needed. Our high cadence, high precision RV observations of the 39 DMPP targets (Figure 1) are described in Methods and Supplementary Information. Supplementary Table 4 summarises the status of the programme.

## Hypothesis Testing

DMPP was predicated on the hypothesis that anomalously low values of log $R'_{HK}$ are due to circumstellar gas ablated from short period planets. This hypothesis predicts the DMPP targets should be hosts of planets with periods of a few days or less. We have collected observations on 17 targets (Supplementary Table 4). Our 100% planet detection rate where we have sufficient measurements, qualitatively implies the DMPP selection criteria indeed pick out short period planet hosts.

To quantify the statistical evidence, we use planeticity statistics derived from Kepler. The probabilities of a single star hosting multiple low mass, close-in planets are not independent: the prevalence of compact multiplanet systems is due to evolutionary processes which create multiple low mass planets in short period orbits around significant subset of stars. To assess the probability of randomly choosing DMPP-1, the host of such a system, we consider how many of them were discovered by Kepler. There are more than 439 Kepler transiting multiplanet systems, i.e., around 4 per 1000 stars observed (**https://exoplanetarchive.ipac.caltech.edu**). The majority of systems will not transit, so compact multiplanet systems are significantly more common than this. To estimate the prevalence of compact multiplanet systems similar to DMPP-1, we use Kepler main mission discoveries of systems with two or more transiting planets with periods shorter than 5.52 d, corresponding to DMPP-1e, the second innermost detected planet orbiting DMPP-1. Kepler measured $N_{KM} = 43$ such systems. Assuming a random orientation, the transit probability of DMPP-1e is $p_T = 0.09$, so the discovery of $N_{KM}$ systems among the $N_{KO} = 100,000$ stars Kepler observed implies $f_{CM}$, the fraction of stars hosting compact planetary systems akin to DMPP-1 is

$$f_{CM} \approx \frac{N_{KM}}{N_{KO}} \times \frac{1}{p_T} \approx \frac{43}{100,000} \times 11 \approx 0.005 \;.$$

The occurrence of systems like DMPP-2 is[54] $f_{HJ} = 0.0025 \pm 0.0005$. DMPP-3 is unique so it is very difficult to quantify its rarity. We can derive a firm upper limit on the relevant occurrence rate from the occurrence of super-Earths with parameters resembling DMPP-3Ab, ignoring the fact that membership of an eccentric compact binary star system makes this a much less likely configuration; here we expect[54] $f_{SE} = 0.053 \pm 0.01$. This uses the probabilities for planets of the appropriate category existing in orbits of period < 5.9 d and < 10 d for DMPP-2b and DMPP-3Ab respectively.

The probability, $p$, of discovering systems like DMPP-1,-2,-3 in an unbiased sample of 17 targets using the multinomial distribution is:
$$p = \frac{17!}{1! \times 1! \times 1! \times 14!} \, f_{SE} f_{CM} f_{HJ} f_0^{14} = 0.0011 \;,$$

where $f_0 = 1 - f_{SE} - f_{HJ} - f_{CM}$ is the probability of a null result. The probability that we would have made the three DMPP planetary system discoveries in a random sample of 17 main sequence stars is around 0.1%. There are possible alternative evaluations of $p$: for example, if we used $f_{HJ} = 0.012$, the RV-based probability of a randomly selected star hosting a hot Jupiter (P < 10 d, $M_P \sin i > 0.1$ $M_J$), we would obtain[55] $p = 0.0047$ which is still very small.

This calculation grossly underestimates the rarity of DMPP-3, and ignores the fact that, far from being null results, the remaining 14 targets are either under-observed or constitute preliminary discoveries of close-in planetary systems. A conservatively estimated one in a thousand chance of obtaining our results randomly implies that the hypothesis which motivated DMPP is validated. Our sample of main sequence stars with log $R'_{HK}$ < -5.1 appears to be strongly biased towards the presence of short-period planets. We conclude the Ca II H&K chromospheric line core deficits are associated with the presence of short period planets, probably via absorption by circumstellar gas ablated from them.

The alternative or additional hypothesis, attractive for the elegance with which it explains the upper and lower boundaries of the Neptunian desert[51], is that circumstellar gas shrouds may be remnants of tidally disrupted planets. This tidal hypothesis predicts the presence of longer period companions capable of scattering planets on to eccentric orbits with a small periastron distance from the star. The 17 observed objects in Supplementary Table 4 include two which fit this hypothesis: DMPP-3 and candidate LP-S. This is encouraging, but as we have not yet performed observations and analysis designed to detect long period companions it would be premature to attempt quantitative tests of this hypothesis, particularly as the Galactic demographics of dim, massive, long-period companions are uncertain.

# Conclusions and Future Outlook

DMPP is a very efficient planet search. It is detecting low mass planets orbiting bright, nearby stars, which are particularly amenable to compositional studies of the planetary surface. Since the DMPP targets have ablated planetary material filling our line of sight to the chromospherically active regions of the host star, angular momentum considerations favour approximately edge-on viewing angles. The transit probability for DMPP systems exceeds that of randomly oriented orbits by a factor dependent on how closely the ablated material is confined around the orbital plane. Transits of rocky planets are, in general, too shallow to be detectable from the ground, so we need space-based photometry to search for transits. As our targets are bright, nearby stars they will be observed by TESS. This will lead to statistical constraints on the geometry of the circumstellar gas, in turn guiding simulations of mass loss from close in planets[8]. High quality mass and radius measurements are possible for the transiting DMPP planets. Assuming the circumstellar gas shrouds are not perfectly azimuthally symmetric, the DMPP systems are also amenable to transmission spectroscopy techniques to determine the elemental abundances of the ablating rocky surfaces.

Kepler 1520b is the prototype of exoplanets showing the most dramatic short timescale planetary mass loss. It was discovered through the variable-depth transits of a dust cloud condensing from the ablating gas[14]. In Kepler 1520b and similar objects, the metal-rich gas which co-exists with the dust will undoubtedly absorb strongly in Ca II H&K. These Kepler systems are too faint for reliable measurements of log $R'_{HK}$ and for high-precision RV work; so the DMPP survey was partly motivated by the aim of finding nearby bright analogues and progenitors of them. The DMPP systems may thus host objects similar to Kepler 1520b on orbits interior to our RV-detected planets. Hence TESS and CHEOPS observations may reveal transiting dust clouds, or analogous non-transiting systems through forward scattering of starlight[56].

This potentially allows us to extend exogeology to dozens of objects beyond the Solar System. DMPP may provide key examples for comparative planetology in the 2020s.

**Table 1:** Maximum likelihood and *a posteriori* Keplerian parameters for DMPP-2. The solution using the combined HARPS data is shown in column 1, while the solution with HARPS data treated as two data sets is listed in column 2. The combined HARPS+MIKE data set solutions with eccentricity and without eccentricity ($e = 0$) are respectively shown in columns 3 and 4. Rows 1 to 2 are the maximum likelihood and false alarm probabilities. Rows 3 to 9 list the *a posteriori* parameters for each model. $\lambda = M_0 + \omega$ is the mean longitude at the reference epoch $t_0$. The derived parameters are listed in rows 10-12 with further information about the data sets in rows 13-15. Parameter posterior uncertainties and derived uncertainties are shown in parentheses. See DMPP-1 article (*this issue of NA*) and Methods section of Anglada-Escude et al.[20] for further details of likelihood modelling and error determination from posterior parameter deternination.

| Parameter | HARPS ($\gamma_{HARPS} = \gamma_{P95+P97+P98}$) | HARPS ($\gamma_{P95+P97}, \gamma_{P98}$) | MIKE+HARPS ($\gamma_{MIKE}, \gamma_{P95+P97}, \gamma_{P98}$) | MIKE+HARPS ($\gamma_{MIKE}, \gamma_{P95+P97}, \gamma_{P98}$) $e = 0$ |
|---|---|---|---|---|
| **Maximum likelihood parameters** | | | | |
| FAP | $5.9 \times 10^{-7}$ | $9.1 \times 10^{-9}$ | $2.0 \times 10^{-7}$ | $8.4 \times 10^{-8}$ |
| $\Delta \log L$ | 25.5 | 31.5 | 31.3 | 32.3 |
| **Maximum *a posteriori* parameters** | | | | |
| $\chi_r^2$ | 1.000 | 1.006 | 1.005 | 1.154 |
| $P$ [d] | 5.9981 (5.9952 - 6.0002) | 5.2049 (5.2028 - 5.2064) | 5.2072 (5.2017 - 5.2074) | 5.2070 (5.2018 - 5.2074) |
| $K$ [ms$^{-1}$] | 41.875 (35.71 - 45.15) | 39.99 (35.79 - 43.62) | 40.26 (34.86 - 42.95) | 39.13 (34.49 - 42.57) |
| $e$ | 0.022 (< 0.067) | 0.083 (< 0.082) | 0.078 (< 0.082) | 0 |
| $\lambda = M_0 + \omega$ [deg] | 30.11 (22.27 - 38.96) | 13.45 (6.36 – 20.99) | 271.54 (232.57 – 290.33) | 259.03 (229.48 – 291.79) |
| $\gamma_{HARPS}$ [ms$^{-1}$] | -5.63 (-8.77 - -3.03) | | | |
| $\gamma_{MIKE}$ [ms$^{-1}$] | | | 14.68 (-5.74 - 26.02) | 19.47 (-5.15 - 26.31) |
| $\gamma_{P95+P97}$ [ms$^{-1}$] | | -14.84 (-17.40 - -11.50) | -14.48 (-16.78 - -10.74) | -13.47 (-16.42 - -10.42) |
| $\gamma_{P98}$ [ms$^{-1}$] | | 7.10 (2.69 - 12.00) | 7.35 (2.44 - 11.76) | 8.36 (2.37 - 11.72) |
| $\sigma_{HARPS}$ [ms$^{-1}$] | 18.53 (17.08 - 21.22) | | | |
| $\sigma_{MIKE}$ [ms$^{-1}$] | | | 18.05 (16.69 – 39.83) | 15.81 (15.38 – 38.92) |
| $\sigma_{P95+P97}$ [ms$^{-1}$] | | 12.26 (11.46 - 16.75) | 13.22 (12.73 - 17.06) | 13.99 (13.02 - 17.37) |
| $\sigma_{P98}$ [ms$^{-1}$] | | 18.14 (16.69 - 24.06) | 17.35 (16.81 - 24.35) | 18.30 (16.82 – 24.42) |
| $M_p \sin i$ [M$_J$] | 0.477 (0.405 - 0.516) | 0.434 (0.387 - 0.475) | 0.437 (0.378 - 0.467) | 0.426 (0.374 - 0.464) |
| $a$ [AU] | 0.0730 (0.0724 - 0.0735) | 0.0664 (0.0659 - 0.0668) | 0.0664 (0.0659 - 0.0669) | 0.0665 (0.0659 - 0.0669) |
| $N_{obs}$ | 49 | 49 | 56 | 56 |
| Data baseline [d] | 476 | 476 | 4508 | 4508 |
| $t_0$ [BJD-2400000] | 57286.78871 | 57286.78871 | 53254.84461 | 53254.84461 |

## Acknowledgements

This work is based on observations collected at the European Organisation for Astronomical Research in the Southern Hemisphere under ESO programmes 081.C-0148(A), 088.C-0662(A) and 091.C-0866(C), 095.C-0799(A), 096.C-0876(A), 097.C-0390(B), 098.C-0269(A) 098.C0499(A), 098.C0269(B), 099.C-0798(A) and 0100.C-0836(A). The research leading to these results has received funding from the European Community's Seventh Framework Programme (FP7/2007-2013 and (FP7/2013-2016) under grant agreements number RG226604 and 312430 (OPTICON). These results were based on observations awarded by ESO, OPTICON and OHP using HARPS, HARPS-N and SOPHIE. This research has made use of the SIMBAD data base, operated at CDS, Strasbourg, France. We would like to thank Pamela Arriagada (who was unavailable at the time of article submission) for providing the archival observations from the MIKE spectrograph at the Las Campanas Observatory. D.S. was supported by an STFC studentship. CAH and JRB were supported by STFC Consolidated Grants ST/L000776/1 and ST/P000584/1; DS, JSJ and JC were supported by STFC studentships. GA-E was supported by STFC Consolidated Grant ST/P000592/1; JSJ acknowledges support by FONDECYT grant 1161218 and partial support from CONICYT project Basal AFB-170002. We thank several anonymous referees of DMPP papers for their constructive comments.

## Author Contribution

CAH plans and leads all aspects of the DMPP collaboration, secured the funding, wrote the proposals, and the paper. DS performed target selection and initial RV analyses and contributed to proposal writing. JRB performed final RV analyses, contributed to proposal writing and co-wrote the paper with GA-E providing software and expertise. LF contributed to the analysis and proposal writing. AJN carried out analysis of the SuperWASP photometry data. JSJ provided expertise on stellar activity and the log $R'_{HK}$ metric. CAH, DS, JRB, JSJ and JC performed observations with HARPS and SOPHIE. All authors were given the opportunity to review the results and comment on the manuscript.

## Author Information

Reprints and permissions information is available at www.nature.com/reprints. Correspondence and requests for materials should be addressed to Carole.Haswell@open.ac.uk

## Data availability statement

Correspondence and requests for materials should be addressed to Carole.Haswell@open.ac.uk

## Main text figures

**Figure 1:** : The sample of 2716 main sequence stars from which our targets were drawn[11] and known hot Jupiter hosts[12]. The maximum activity, log $R'_{HK}|_{max}$, is plotted for each star. Our targets are the main sequence stars that fall below the basal limit.

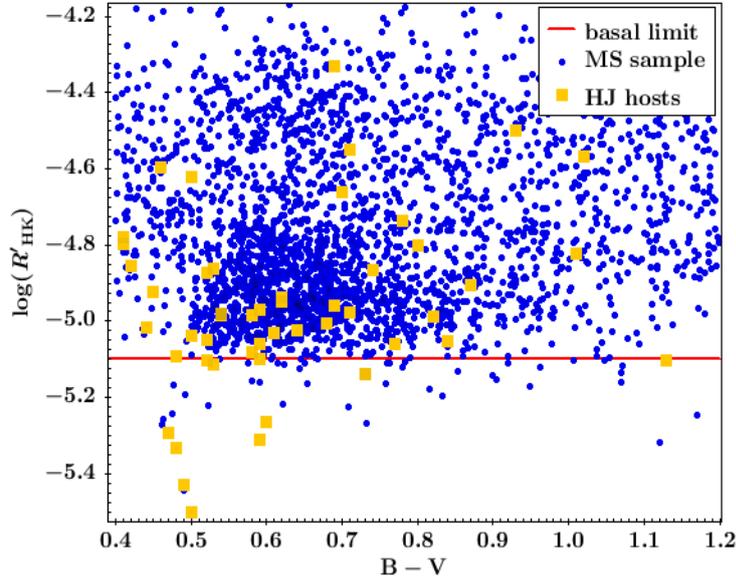

**Figure 2:** Distance and magnitude distributions of DMPP targets. Upper panel: Distances of the DMPP targets, compared to distances of the parent main sequence sample. Lower panel: Distances and apparent V magnitudes for the parent main sequence sample (blue), known transiting planet hosts (green) and known RV planet hosts.

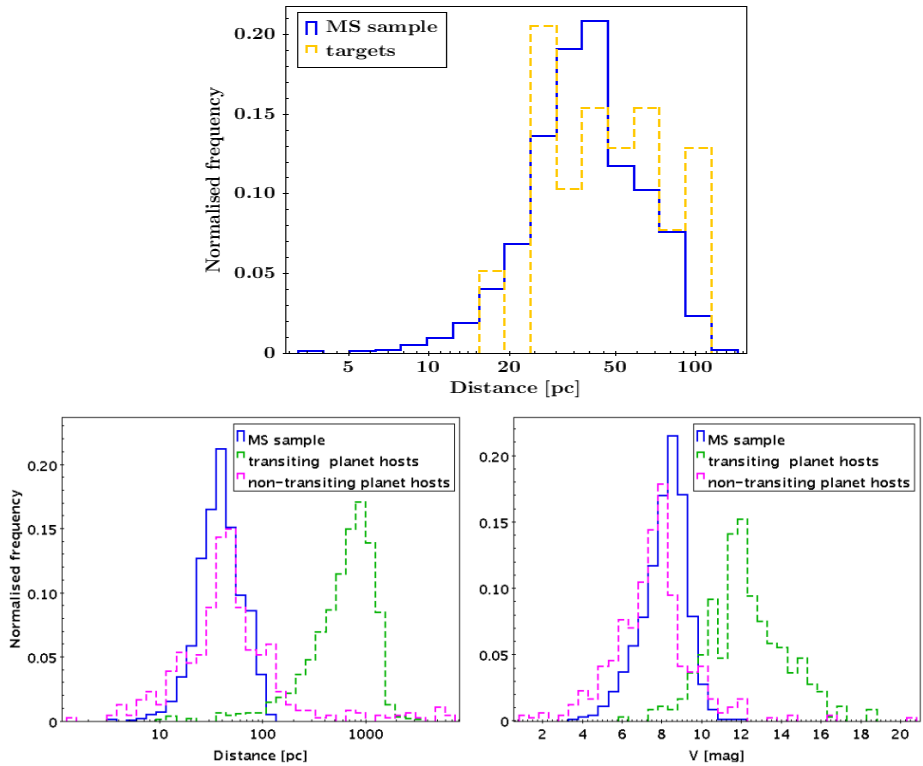

**Figure 3:** Observations of DMPP-2 with MIKE and HARPS showing derived radial velocities and corresponding uncertainties (RV error bars are 1-σ throughout). The maximum *a posteriori* model with systemic offsets of $\gamma_{MIKE}$, $\gamma_{P95+P97}$ and $\gamma_{P98}$ (Table 2, column 3) is plotted.

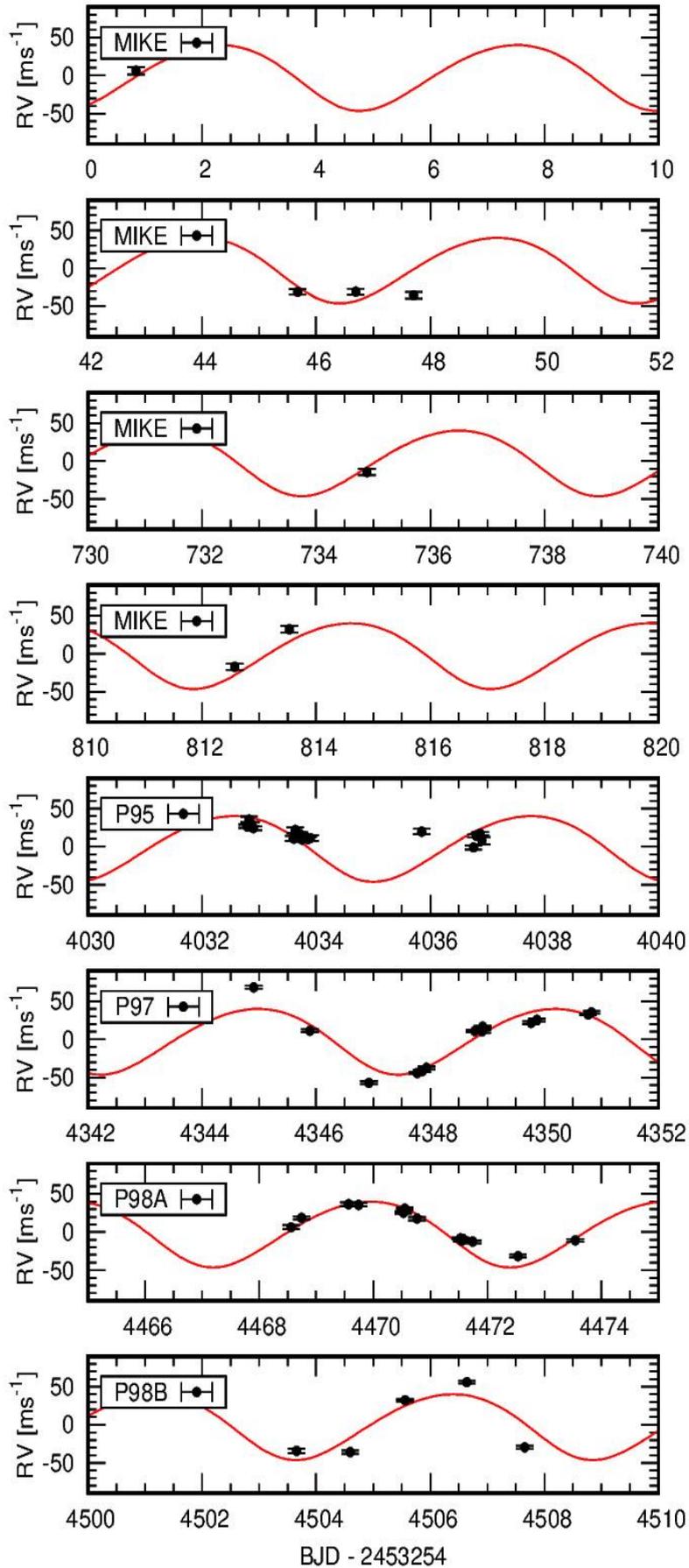

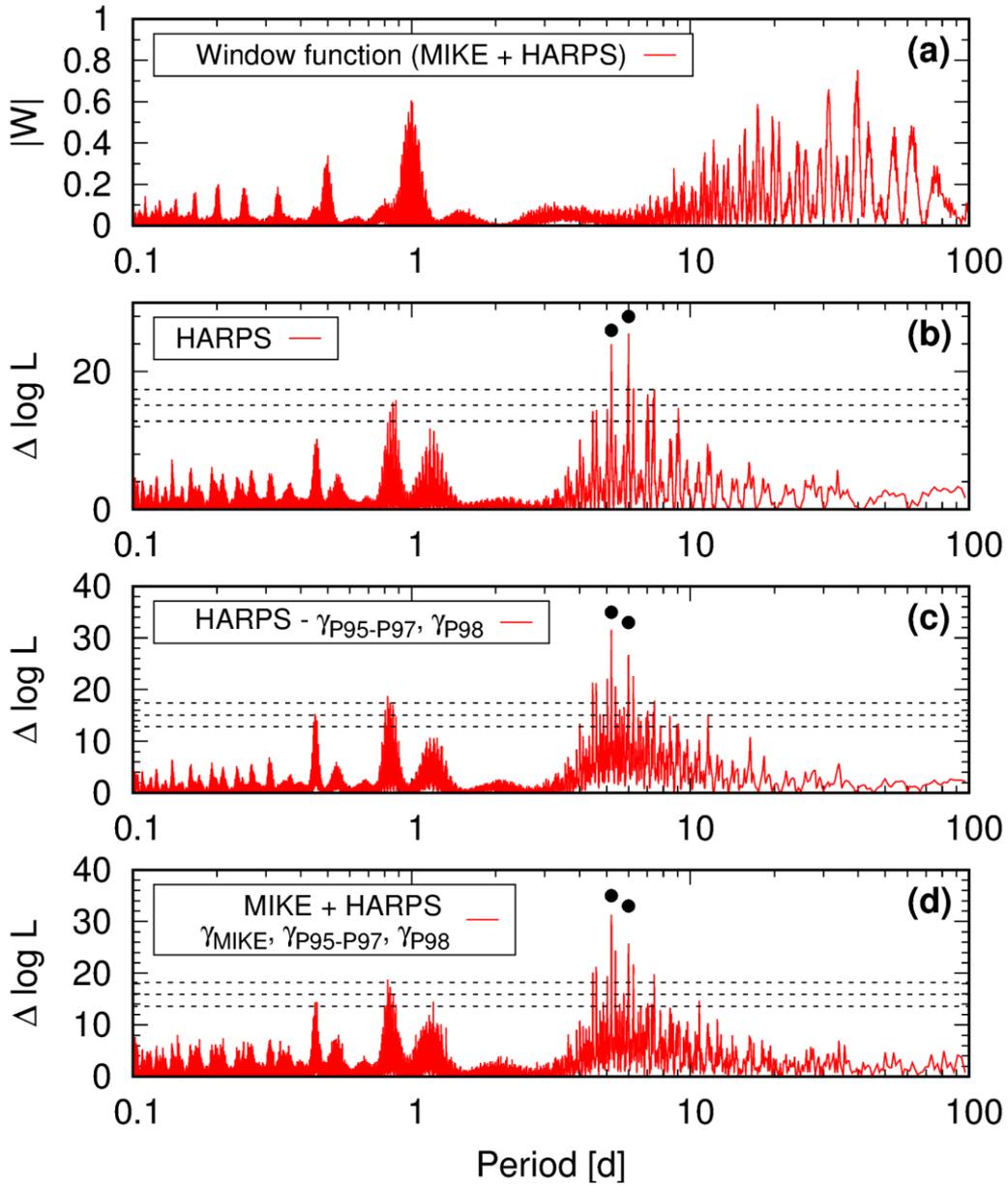

**Figure 4:** Window function and log-likelihood periodograms for DMPP-2 radial velocities. Panels show (a) the window function, (b) the combined HARPS data, (c) the HARPS data when fitting for a zero-point offset between the P95-P97 and the P98 subsets and (d) the MIKE + HARPS with zero-point offsets. For each periodogram the $P_{orb} = 5.998$ d peak is indicated, with peaks at $P_{orb} = 5.206$ d, 5.205 d & 5.207 d respectively for panels b, c & d. The 10, 1 and 0.1% FAP thresholds are shown.

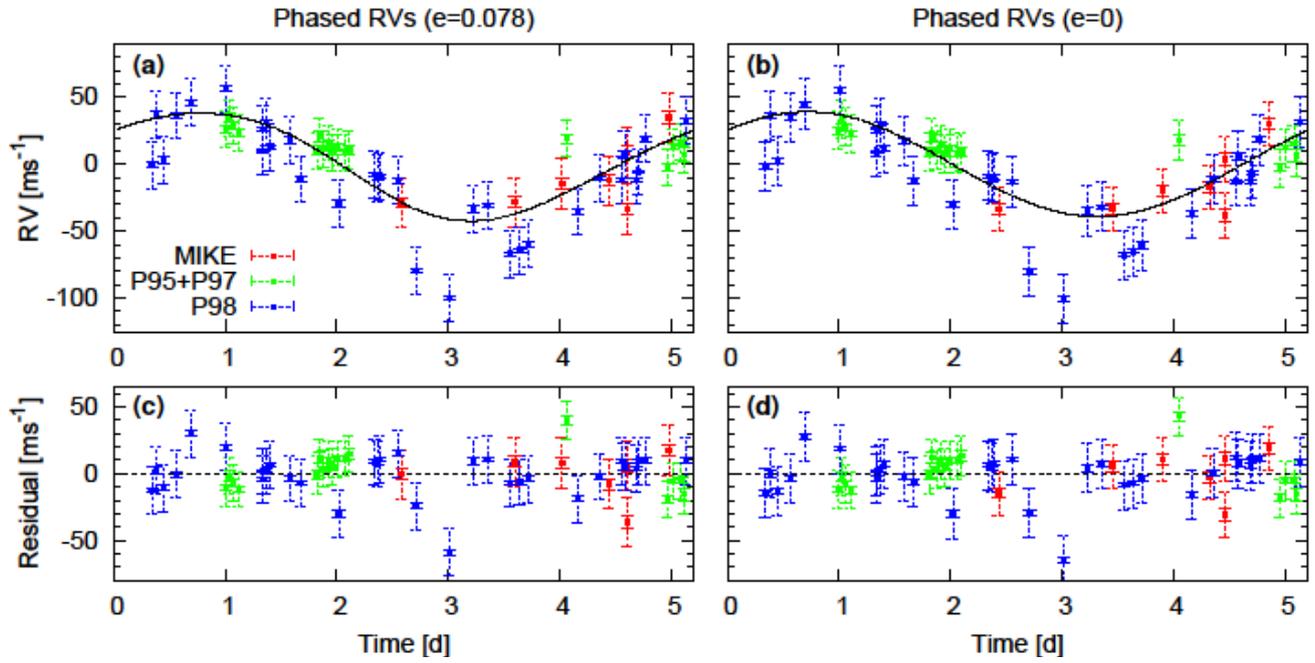

**Figure 5:** Folded radial velocities for DMPP-2 corresponding to the maximum *a posteriori* solutions in Table 2. The combined solutions with MIKE and HARPS observations for (a) $e = 0.078$ and (b) $e = 0$ are shown. The respective residuals are shown in panels (c) and (d). The observation uncertainties are shown as thick error bars while the thinner dotted error bars are the uncertainties with the appropriate jitter, $\sigma_{MIKE}$, $\sigma_{P95+P97}$ and $\sigma_{P98}$, from columns 3 and 4 of Table 2.

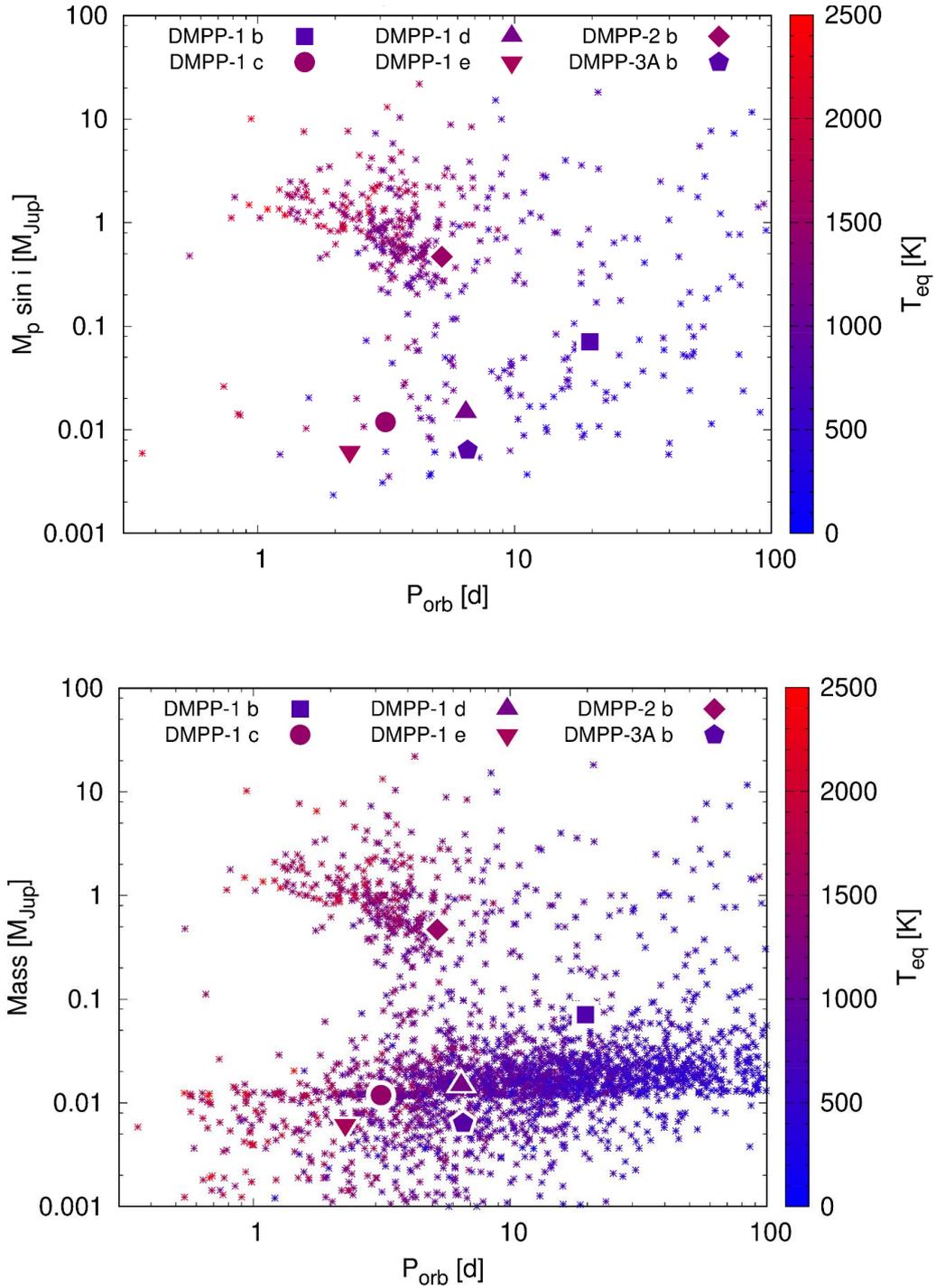

**Figure 6:** $M_p \sin i$ vs orbital period for the first three DMPP systems. Our DMPP planets are plotted with (a) other planets with RV mass determinations and (b) with the addition of transiting planets with model-dependent masses derived from radius measurements.

## Methods

### Target Selection

We begin[57] with 7864 stars[11] and used the Extended Hipparcos Compilation[58] to exclude blended objects, multiple stars, objects without reliable *B-V* and $M_V$ values, and stars outside the range 0.4 < *B-V* < 1.2, corresponding to spectral types F4 - K5. The latter constraint arises because log $R'_{HK}$ is only well-calibrated for FGK stars. 5004 stars remained. We carefully examined the primary data reported[11] and recalculated log $R'_{HK}$ from the primary literature S-values[57,59,60]. The log $R'_{HK}$ values in the literature suffer from systematic calibration offsets, and are often reported without error bars[9]. The error budget is particularly important for stars with low values of log $R'_{HK}$, as their Ca II H&K line cores have very little flux. Stellar activity is intrinsically variable on a variety of timescales, and our goal was to pick the stars most likely to be affected by circumstellar gas shrouds, so we used the maximum values of log $R'_{HK}$ for each star to select our targets[57]. Our sample satisfy log $R'_{HK}|_{max}$ < -5.1. After our sample selection was completed, subsequent measurements of log $R'_{HK}$ > -5.1 can arise. Indeed, after we completed our RV observations, log $R'_{HK}$ = -5.056 was reported[61] for DMPP-2, though we suspect there may be some calibration issues in these log $R'_{HK}$ values. The log $R'_{HK}$ value in (Supplementary Table 1) is robust, being a median of 7 values spanning over 3 years, each below -5.1 and ranging between -5.36 and -5.17 [P. Arriagada, private communication]. Any single measurement depends on the (presumably variable) intrinsic stellar activity and the (presumably variable) column of absorbing circumstellar gas.

The basal level of chromospheric emission, log $R'_{HK}$ = -5.1, only applies to main sequence stars, so we retained only 2716 stars with luminosities within $\Delta M_V$ = 0.45 of the empirical average main sequence[62,57] which corresponds to stars with a metallicity of [Fe/H] ~0.3. This limit was chosen as a compromise between retaining metal-rich, unevolved stars and rejecting metal-poor, evolved stars. Our sample therefore has biases dependent on stellar metallicity, but these are not straightforward to quantify. This is because, in addition to the metallicity-dependent main-sequence luminosity, the stellar Ca II H&K (photospheric and chromospheric) line profiles will be metallicity-dependent, and the circumstellar absorption depressing the line cores depends on the Ca abundance in the ablated planetary gas.

Our sample has distances below ~100 pc and 5 ≤ V ≤ 11, i.e. bright, nearby stars (Fig 2). We found 39 stars with log $R'_{HK}|_{max}$ < -5.1. For each of these we calculated lower limits for the interstellar (IS) column required to produce their Ca II H&K line core deficits[13], finding $N_{CaII} \geq 10^{12.5}$ - $10^{14}$ cm$^{-2}$ is required. This is a factor of 100 larger than the average IS column at these distances, and larger than the column measured along any of a sample of 266 lines of sight[63]. It is possible some of our targets are selected due to anomalously high interstellar absorption, but it would be no easier to ascertain this than to search for the putative low mass ablating planets.

Sometimes observability constraints on the allocated nights required us to consider auxiliary targets. Consequently, we have collected RV data on seven targets outside the initial sample of 39 discussed above. Two stars satisfy log $R'_{HK}|_{mean}$ < -5.1, while five appear to have depressed log $R'_{HK}$ for their evolutionary context. For clarity and conciseness, we have excluded them from the statistical analysis and hypothesis testing reported herein; two of them are promising compact multiplanet system candidates.

### Observing Strategy and Data Reduction

We suspect our target stars host low-mass ablating planets on short-period orbits, consequently we designed a programme of high cadence, high precision RV measurements. We carried these out at the European Southern Observatory (ESO) using HARPS and at Observatoire de Haute Provence (OHP) using SOPHIE for Northern Hemisphere Targets, we have also observed a handful of objects on HARPS-N. Our SOPHIE and HARPS-N observations were awarded via OPTICON. To be sensitive to short orbital periods, we observed each star several times per night, cycling between 2-5 target stars and an additional RV standard star. We exposed for at least 15 min, to average over asteroseismic oscillations. This generally gave a signal to noise ratio (SNR) exceeding 100 at 550 nm, and during poor conditions we increased the exposure time to achieve this but imposed a maximum exposure time of 40 min to limit uncertainties due to the changing barycentric velocity of Earth during exposures with variable transparency and seeing. Since we were observing in visitor mode, our data are grouped in intervals of intensive coverage interspersed by gaps often ~ 1 year. This made us sensitive to periodic RV signals of hours to days, but determining the genuine signal from two or more aliases often proved difficult. We cannot bin data from exposures separated by hours to days and remain sensitive to stellar RVs on these timescales. Consequently, though we often achieved measurement errors of < 1 m s$^{-1}$, most of our observations will suffer from granulation noise[64] of > 1 m s$^{-1}$.

Basic data reduction and the calculation of RVs via the cross-correlation technique was performed at the telescopes by the Data Reduction Suite (DRS) [Bouchy, F. and Queloz, D. *HARPS DRS USER Manual* www.eso.org/sci/facilities/lasilla/instruments/harps/doc/DRS.pdf], which also calculates the full width at half maximum (FWHM) of the cross-correlation function (CCF) and the bisector span[65] (BIS). We used much of the DRS output, but refined the RV determination using HARPS-TERRA[66] which produces improvements to RV precision by using more of the echelle orders and using the stellar spectrum as the cross-correlation template. RV precision was key to our project, we carefully examined and corrected for CTI effects, and examined our data for the chromatic effect[57,67].

## CTI correction

Precision RVs require measurements of line centroid shifts corresponding to ~ 0.001 pixels, so charge transfer inefficiency[68] (CTI) can be a significant problem. CTI causes systematic changes to the stellar line profiles which are correlated with the charge accumulated. This leads to a spurious dependence of the measured RV on the SNR.

The HARPS-N DRS automatically corrects for its low-level CTI effect[69], but the HARPS DRS does not (Lovis, private communication). For M dwarfs observed with HARPS an empirical correction[70] is

$$\frac{\Delta RV_{CTI}}{[\text{m s}^{-1}]} = 4.92 - 1.31 \ln(SN_{60}), \qquad \textbf{Equation 1}$$

where $SN_{60}$ is the SNR at order 60. This correction removes a spurious ~ 2 m s$^{-1}$ red-shift at low SNR. Our targets have a different flux distribution and weighting of lines in the CCF, so we used archival data to derive the corresponding correction for FGK stars.

We used observations of HD1581, HD85512, HD154577 and HD190248 taken before the HARPS instrumental upgrade with $30 < SN_{60} < 500$. We applied offsets to each of these RV standard stars, shifting the HARPS-TERRA RVs so the median value of RV for observations with $100 < SN_{60} < 130$ were the same for each star (Supplementary Fig 1). Retaining the functional form of Equation 1, the nonlinear least-squares fit to the relationship between RV and $SN_{60}$ is

$$\frac{\Delta RV_{CTI}}{[\text{m s}^{-1}]} = 4.88 - 1.03 \ln(SN_{60}). \qquad \textbf{Equation 2}$$

Supplementary Fig 1 shows there is little difference between the two corrections. We also fitted each RV standard separately, the resulting correction laws agreed within error. No clear dependence of the CTI on stellar spectral type was seen.

DMPP HARPS data have ~50 < $SN_{60}$ < ~120 corresponding to differential corrections up to 0.9 m s$^{-1}$. In practise, the CTI correction was not an important effect in our HARPS observations.

For SOPHIE, CTI causes a much larger systematic offset, reaching ~10 m s$^{-1}$ at $SNR$ ~ 70 and ~20 m s$^{-1}$ at $SNR$ ~40. We used the SOPHIE flux-meter to facilitate consistent SNR for several targets, but this was sometimes precluded by variable conditions at OHP and requiring exposure times ≤ 40 mins. CTI effect correction is consequently vital, and has not been published for our $SNR$ range. We derived a power-law correction[70,71] from publicly available SOPHIE+/HR mode data on HD 221354, HD 30708, HD 185144, HD 139324, HD 109358 and HD 5372 with 17 < $SNR$ < 260. After offsetting data on each star to align the median RV at SNR > 140, a power law was fit using non-linear least squares (Supplementary Fig 2):

$$\frac{\Delta RV_{CTI}}{[\text{m s}^{-1}]} = -3170\, SNR^{-1.37} + 2. \qquad \textbf{Equation 3}$$

As shown in Supplementary Fig 2, the updated relation is a significant improvement. Additional observations with $SNR$ < 60 would allow further improvement. Applying Equation 3 to our targets' SOPHIE RV timeseries led to differential RV corrections of up to 10 m s$^{-1}$.

## The Chromatic Systematic Effect

Our observing strategy requires observations at a range of airmasses, potentially exposing us to a serious systematic effect. Differential atmospheric refraction causes colour-dependent slit losses. This is ameliorated by the use of an atmospheric dispersion corrector, but the correction can be imperfect[67,72]. This leads to an airmass-dependent apparent change in the colour of a target as illustrated in Supplementary Fig 3. In turn, this changes the weighting of the stellar spectral lines in the CCF used for RV measurement, thus potentially causing erroneous RV signals. To examine this effect, we characterise the pseudo-spectral energy distribution (pSED) with slope $\kappa$ determined empirically from a linear fit to the normalised pSED within the wavelength range shown in Supplementary Fig 3 following the procedure outlined for observations made with HARPS-N[72].

The DRS installed at HARPS and HARPS-N corrects for the chromatic effect[69] and we have found no evidence for RV-$\kappa$ correlations in our data from these instruments. In contrast, for SOPHIE this correction is not implemented, and we must account for this systematic effect in our analysis by including an RV-$\kappa$ correlation in our log-likelihood period analysis (see "RV Signal Detection" in Methods and "Examining correlations between DMPP-2 RVs and Stellar Activity Indicators" below).

## RV Signal Detection

We adopt a frequentist and Bayesian approach[20], with the likelihood function defined to allow global (simultaneous) optimisation of planetary signals, noise parameters (jitter), and linear correlations between RVs and other simultaneous observables. Since the parameters describing multiple planet signals and / or RV-activity index trends are correlated, this approach is required for reliable estimation of the significance of planetary signals[66,74].

We use likelihood ratio periodograms, which show the improvement of the likelihood statistic, $\Delta \log L$, of a best fit including a planetary signal compared to the null hypothesis which is the best fit without a Keplerian signal. We generally treat data from different DMPP observing runs as distinct subsets[74]. This accommodates poorly-constrained possible long-period planetary or stellar signals and allows for different jitter values for individual runs. The latter may be necessary if a target star shifts from lower to higher activity phases during the relatively long gap between runs.

In the first instance we generally set the eccentricity of all Keplerians signals to zero. For the short period orbits of prime interest, tidal circularisation is expected to significantly dampen orbital eccentricities over the systems' lifetime[75]. However, gravitational interactions may drive eccentricity for some of our ablating planets such as DMPP-3Ab so we investigate by comparing maximum likelihood solutions with and without forcing zero eccentricity. A similar iterative investigative approach is adopted in our exploration of correlations between activity indicators and RVs for all targets and in assessing the chromatic systematic effect.

There is no universally agreed value in the literature on the threshold for claiming statistically significant planet detections[76]. False alarm probabilities (FAP) of 1% and 0.1% are typically used. We generally adopt the latter, but report additional less significant signals in each system. We define the *a posteriori* 68.3 per cent confidence intervals for our detected signals through Markov chain Monte Carlo (MCMC) samplings with priors chosen to be uniform and uninformative except orbital eccentricity, $e$, which assumes a Gaussian with zero mean, defined for $e \geq 0$, *i.e.*, with $P(e) = \exp(-e^2/2\sigma_e^2)$; we set $\sigma_e = 0.05$ for DMPP-2. See DMPP-1 article (*this NA issue*) and Methods section of Anglada-Escude et al.[20] for further details of likelihood modelling and error determination from posterior parameter deterination.

## DMPP Observational Status Report

Supplementary Table 4 summarises our observations and analysis. One star has been dropped as analysis of the DMPP spectra revealed log $R'_{HK}$ > -5.1. For this star, the archive values[11] may be erroneous, but it is possible the discrepancy is astrophysical. In either case, this star no longer satisfies our selection criterion log $R'_{HK}|_{max}$ < -5.1 (see Methods), leaving us with 38 targets.

The low mass planets DMPP seeks produce reflex RV amplitudes comparable to the RV precision of state-of-the art spectrographs. Consequently, DMPP planet discoveries generally require more than 60 RV measurements, with disproportionate returns with several consecutive nights of high quality, high cadence observations. DMPP has gathered more data satisfying this constraint in the south than in the north: DMPP-1,-2,-3 and candidate P1-S (which has a preliminary designation *DMPP-5*) host planets discovered using multi-run, high cadence HARPS data.

The northern target P1-N (preliminary designation *DMPP-4*) is a very probable compact multiplanet system, but most of our observations were made with SOPHIE and the characterisation of multiple low amplitude signals is hampered by RV uncertainties comparable to the reflex RV amplitudes we seek. With the inclusion of HARPS-N data, P1-N shows low amplitude 2.54d and 0.82d signals, with some evidence for a third 1.66d signal.

P1-S (*DMPP-5*) is probably also a compact multiplanet system. It exhibits signals indicating up to three planets: two short period super-Earths ($P_{orb}$ = 3.6 d, 6.4 d; $M_P \sin i$ = 3.5 $M_\oplus$, 3.8 $M_\oplus$ respectively); more data is needed to explore a formally very significant ~90 d, $M_P \sin i$ ~ 0.15 $M_J$ signal.

Invariably where we have included more than 60 observations in our analysis, we have evidence for planets. For candidate LP-S, we have only 6 DMPP observations but archival observations reveal a long period 4 $M_J$ planet. We de-prioritised two targets when our early observations appeared to exclude planets more massive than 2 $M_\oplus$ in sub-day orbits. However, the RV curve of DMPP-1 shows intervals where the RV variability is below our detection threshold. We consequently now consider these two targets worthy of further observations. DMPP is an effective and efficient low mass planet search.

## RV Detection Limits

Supplementary Fig 9 illustrates the detection limits we can achieve with our high cadence, high precision RV observations. In this particular case, the simulation assumes 94 HARPS observations made over a 3 yr baseline in six different runs plus single ESPRESSO observations on ten consecutive nights. This is more data than we have currently collected on any DMPP target. The simulations use extant HARPS measurements of one of our targets with a 1.2 $M_\odot$ star and r.m.s. 0.8 m s$^{-1}$ and an assumed ESPRESSO r.m.s. scatter of 0.4 m s$^{-1}$. The addition of ESPRESSO measurements is particularly important to drive down our planet mass detection threshold. With FAP = 0.1% we can get down to a threshold of ~0.5 $M_\oplus$ in very short period orbits, and below 1 $M_\oplus$ for almost all periods ≤ 1.1 d. The DMPP targets are particularly promising stars for observations with GIARPS, ESPRESSO and other state-of-the-art spectrographs capable of sub-m s$^{-1}$ precision.

One of the motivations for DMPP was to seek the nearby analogues and progenitors of planets like Kepler 1520b: ~0.1 $M_\oplus$ rocky bodies in sub-day orbits where they suffer sublimation. The subliming mineral surface is likely to create an extended circumstellar cloud of metal-rich vapour[14]. Even with the latest spectrographs, this is a challenging mass limit to reach, but we hope to detect higher mass, progenitor objects in sub-day orbits.

## Data availability Statement

The data that support the plots within this paper and other findings of this study are available from the corresponding author, Carole.Haswell@open.ac.uk, upon reasonable request.

## Code availability Statement

Requests for codes used in this paper, where they are not publicly available, should be addressed to the corresponding author, Carole.Haswell@open.ac.uk.

## Methods references

# Supplementary Information

## The star DMPP-2

We used the high SNR co-added spectrum from the HARPS observations of DMPP-2 (the F5V star HD 11231) to derive the stellar atmospheric parameters given in Supplementary Table 1[1,2-6]. We computed stellar model atmospheres with the LLmodels code[7] and synthetic spectra with the synth3 stellar atmosphere synthesis code[8] in the same manner as described in references[4-6]. The effective temperature ($T_{eff}$) was obtained by simultaneous analysis of hydrogen (Hα and Hβ) and metal lines (Ti I, Ti II, Cr I, Fe I, Fe II, and Ni I). Synthetic spectra were fitted to the hydrogen line profiles, and excitation equilibrium was imposed for the metallic lines. The surface gravity was derived from fitting the gravity-sensitive Mg I b lines, and imposing ionisation equilibrium for Ti, Cr, and Fe. Synthetic spectra were fitted to 120 weakly blended lines to determine $v\sin i$ and the macroturbulent velocity ($v_{mac}$). The relatively high value of $v_{mac}$ = 8.6 ±1.0 kms$^{-1}$ is not unusual, given the other stellar parameters[9].

Spectral energy distribution analysis of DMPP-2 used synthetic fluxes, calculated with the derived atmospheric parameters, to reproduce the observed Johnson[10] [see Cutri, R.M. et al. "VizieR Online

Data Catalog: 2MASS All-Sky Catalog of Point Sources (Cutri+ 2003)" *VizieR On-line Data Catalog: II/246.* 2246, (2003) and Cutri, R.M. et al. "VizieR Online Data Catalog: WISE All-Sky Data Release (Cutri+ 2012)" *VizieR On-line Data Catalog: II/311* 2311, (2012)] photometry converted to physical units[11-13]. This was combined with the measured parallax (Supplementary Table 1), to estimate the stellar radius ($R_*$) and interstellar reddening, $E(B-V)$. Combining $R_*$ and $T_{eff}$ yields a stellar luminosity $L_* = 1.41 \pm 0.16$ $L_\odot$. Finally, the *param* tool[14] was used for isochrone fitting. The input parameters were the effective temperature, metallicity, V-band magnitude, and parallax, along with a Kroupa mass function[15], assuming a constant star formation rate. This analysis produced the values in the last 5 rows of Supplementary Table 1, and is consistent with the spectral and SED results.

The macroturbulence ($v_{mac}$) and the position of the star in the Hertzsprung-Russell (HR) diagram suggest that DMPP-2 is probably a γ-Doradus star[9]. Thus, non-radial pulsations with periods of the order of ~ 1 d are expected[16,17].

## DMPP-2 Stellar Abundance Analysis

Again, using the coadded HARPS stellar spectrum, we determined the local thermodynamical equilibrium (LTE) abundances of 27 distinct elements (31 ions) from the measured equivalent widths, except for Li I for which the abundance was derived employing synthetic spectra to account for hyperfine structure (HFS)[6]. Supplementary Table 2 lists our measured abundances; solar values[18] are also listed for reference. Where lines of other elements are affected by HFS (e.g. Mn, Cu, Co), we considered just weak, non-saturated lines. This removes the effects of HFS on the derived abundances; we verified this via a lack of correlation between line abundance and line equivalent width. The obtained abundance values are largely in good agreement with Solar, except for a slight overabundance of some Fe-peak elements and of Ba. The latter appears to be a feature of stars similar to DMPP-2[19]; *N.B.* for these stars correcting for non-LTE effects would tend to increase these abundances[20].

## DMPP-2 Archival Radial Velocity Observations

In 2004 – 2006 the Magellan Planet Search Program[21] made 7 observations of DMPP-2 using MIKE. The instrumental stability was ~ 5 m s$^{-1}$ level[22]. This sparsely sampled RV timeseries shows significant variability (Fig. 1): r.m.s. of 23 m s$^{-1}$ and changes of 50 m s$^{-1}$ on a 1 d timescale. This r.m.s. exceeds the internal uncertainties of 4.1 m s$^{-1}$ by more than a factor of 5; adding instrumental RV jitter in quadrature, the uncertainties are 6.5 m s$^{-1}$. The RV variability in Fig. 1 (top panels) motivates high-cadence follow-up; we surmise this was not pursued due to the variable stellar line profiles, see Supplementary Figs 5 & 6.

## DMPP-2 RV Observations with HARPS

We acquired 49 RV measurements of HD 11231 (DMPP-2) over four observing runs with HARPS. These comprised 17 observations in ESO Period P95, 13 in P97, 13 in P98(A) and 6 in P98(B). High cadence observations in P95 showed RV variability of at least ~20 m s$^{-1}$ (peak-to-peak) on timescales > 1 d (Fig. 3). Weather losses in P95 prevented robust detection of periodicities. Subsequently we observed DMPP-2 with 1-3 RV measurements per night, matching the timescale of RV variability. With a time-share agreement in P97 we followed DMPP-2 over a 7 night baseline. We continued observations throughout P98, to confirm the RV periodicity detected in the P95+P97 data and to monitor the prominent stellar line profile changes. Initially exposure times were varied between 900 and 1800 secs, depending on weather conditions. Later in 3 instances we used 600 sec exposures: the increase in stellar jitter contribution from p-mode oscillations is negligible compared to the ~150 m

s$^{-1}$ peak-to-peak variability. Variations in extinction and seeing across our observing runs resulted in a wider range of SNR (50-160) and $\sigma_{RV}$ (1 - 4 m s$^{-1}$, with a median of 2.0 m s$^{-1}$) than usual for DMPP. For DMPP-2, the RV precision is affected by macroturbulent line broadening (Supplementary Table 1). As with DMPP-1 (HD 38677), we used our charge transfer efficiency calibration to correct for the variable SNR.

## Examining correlations between DMPP-2 RVs and Stellar Activity Indicators

Supplementary Fig 4 shows that the BIS and FWHM vary on timescales similar to but distinct from the RV period. To generate the FWHM periodogram (panel c) using the `RECPER` signal search software, our likelihood model included an offset between the P95-P97 and P98 subsets to allow sensitive exploration of short-period variability. Without this the long-term, high-amplitude FWHM shift apparent in Supplementary Fig. 5 has an obscuring effect. The cluster of BIS periodogram peaks around 10 d and the FWHM peaks around 4 d are absent from the RV and window function periodograms. Importantly, there are no significant peaks (<10% FAP) in the S-index periodogram in Supplementary Fig 4 (d). We attribute the periodic variability in the line profiles to γ-Doradus pulsations.

In Supplementary Fig 5 we show BIS, FWHM and S-index against RV along with BIS against RV residuals. Panels a, c & d show that there is no clear correlation between the activity indicators and the RVs; the Pearson's $r$ coefficients are -0.19, 0.13, 0.05 for BIS-RV, FWHM-RV, S-index-RV respectively.

The corresponding $p$ values derived from the F-test, are 0.37, 0.19 and 0.73, implying no significant correlation is detected. The BIS and RV measurements have a very similar amplitude, while the 510 m s$^{-1}$ peak-to-peak FWHM variability is dominated by the offset in P98. The peak-to-peak amplitude ratio BIS/RV is 1.32, and the r.m.s. of the FWHM and BIS is 174 m s$^{-1}$ and 44 m s$^{-1}$ respectively. This CCF parameter variability is rarely seen for inactive, slowly rotating FGK-type stars: in a large HARPS survey of early-type stars, only 3 out of 19 stars with similar basic parameters (0.3 < $B$-$V$ < 0.55 and $v$sin$i$ < 10 km s$^{-1}$) showed a BIS r.m.s. comparable to that of DMPP-2[23]. Among these, HD 138763 has log($R'_{HK}$) = -4.4 indicating high activity. In marked contrast to DMPP-2, HD 138763 shows a clear BIS-RV anti-correlation almost certainly explained by the rotation of active regions[23,24].

We tested RV signal recovery with the inclusion of an activity correlation term using either the BIS or the FWHM timeseries. The activity data, $\xi_i$. are included in the likelihood model as a slope, $C_i$, such that for each data subset, $A_{i,subset} = C_{i,subset} \xi_i$. For the FWHM case, we again treated the P95-P97 and P98 RV data as separate subsets, while we considered both the combined and separate subset cases for the BIS. Although the significance of the detected periodicities (Supplementary Table 3) are reduced slightly, the RV periodograms and Keplerian parameters are essentially unaffected. This confirms the strongest periodicity in Table 1 as a genuine Keplerian signal, robust to removal of correlations with line profile variability.

In general, phase coherence of a stellar (activity or pulsation) signal would be surprising over the 1.3 yr baseline of our HARPS observations. Furthermore, the MIKE observations establish coherence over 12.4 years. The high amplitude and relatively short period of the signal described above might conceivably be explained either by the rotation of a star with prominent and very persistent active regions, or by γ-Doradus pulsations. However, there are no significant correlations between the RVs and either the FWHM or S-index. This rules out the possibility that the RV period is an artefact of stellar variability. There is clearly some periodic line profile variability which produces peaks seen in Supplementary Fig 4. Crucially these periods are distinct from the RV period.

## Searching for Further Keplerian signals in DMPP-2

The search for a second Keplerian signal in the recursive likelihood periodogram framework revealed signals corresponding to the alias peaks in the solution with one Keplerian. For the HARPS data with a single $\gamma_{HARPS}$, the second signal has $P_{orb}$ = 5.211 d with FAP of 0.02. We still find significant jitter of $\sigma$ = 13.0 ms$^{-1}$ with $\chi_r^2$ = 0.998 (cf. 17.8 ms$^{-1}$ and $\chi_r^2$ = 1.000 for the one-Keplerian maximum likelihood solution). With the independent offsets $\gamma_{P95+P97}$ and $\gamma_{P98}$, a second signal with $P_{orb}$ = 7.05 d and FAP of 0.007 is found. The corresponding jitter values for the two subsets are $\sigma$ = 6.4 m s$^{-1}$ and 18.7 m s$^{-1}$ (mean 12.9 m s$^{-1}$) and $\chi_r^2$ = 0.996; this is not a significant improvement on the corresponding one-Keplerian maximum likelihood solution with jitter values of 13.5 and 17.7 m s$^{-1}$ (mean 15.5 ms$^{-1}$) and $\chi_r^2$ = 1.006). We conclude that adding a second Keplerian signal does not significantly reduce the jitter. Therefore, the data do not provide strong evidence that the side-peaks seen in the periodograms correspond to a second planet. For completeness, we note that a second signal at $P_{orb}$ = 1.991 d is found with 22% false alarm probability in the MIKE + HARPS solution. It is probable that this peak is an alias of the $P_{orb}$ = 5.998 d sub-peak. The $\chi_r^2$ is not improved by including this second signal and the residual r.m.s. is only slightly reduced from 15.8 m s$^{-1}$ for one Keplerian to 13.2 m s$^{-1}$ with two Keplerians.

The significant RV variability (jitter) after fitting the single planet solution is not attributable to a second Keplerian signal. Stellar activity can only cause such large jitter for very active stars, at tension with the anomalously low value of log($R'_{HK}$) for DMPP-2.

## RV jitter, interstellar absorption and stellar rotation in DMPP-2

The stellar rotation period of DMPP-2 is unknown. Our $v$sin$i$ and $R_*$ values (Supplementary Table 1) imply $P_*$ < 19 d. A rotation period matching the 5.2 d (or 6 d) RV signal would require a near pole-on orientation: $i$ ~16º (or ~19º). The ASAS and SuperWASP photometric observations of DMPP-2, comprising 450 and over 60,000 points respectively, reveal no significant periodic photometric variability at periods consistent with stellar rotation.

Empirical relationships[25,26] predict RV jitter value of 2-3 m s$^{-1}$ for DMPP-2's stellar parameters and log($R'_{HK}$). The observed ~15 m s$^{-1}$ stellar jitter suggests instead[27] an active star with log($R'_{HK}$) = -4.6. This corresponds to an age of only ~1.1 Gyr[28], contradicting our isochronal estimate, 2 ± 0.3 Gyr. Interstellar absorption cannot plausibly cause a ~0.9 dex depression in log($R'_{HK}$). Using an ISM absorption correction tool[69] this would imply an exceptionally high column density of $N_{CaII}$ > 10$^{15.3}$ cm$^{-2}$. Within 800 pc, the column number density[30], $N_{CaII}$ < 10$^{13}$ cm$^{-2}$, while DMPP-2 lies[2] within 140 pc. Furthermore, the reddening (Supplementary Table 1) is completely inconsistent with such an over-dense ISM.

To reconcile the measured log($R'_{HK}$) = -5.26 with the observed RV jitter, one could speculatively invoke interstellar and strong circumstellar absorption contributions. Even so, a stellar rotation origin for the RV signal would require an unfeasibly large starspot. The lack of correlation between the RV and BIS (Supplementary Fig 5) strongly suggests that any contribution to the jitter from stellar rotation is minor.

## DMPP-2 line profile variations

The BIS and FWHM are two numbers parameterising the line profile shape. Clearly the stellar spectra contain significantly more line profile information, and the CCF gives a high signal to noise aggregation of the line profiles in each of the spectra. We used the full bisector derived from the CCF to explore the behaviour of the DMPP-2 line profiles (Supplementary Fig 6). The bisectors clearly show parallel shifts of ~100 m s$^{-1}$ in addition to significant changes in shape. Our Keplerian solution removes most of the parallel-shift component (Supplementary Fig 6b). Nevertheless, stars with $v$sin$i$ at or below the instrumental resolution, can mimic both the radial velocity variation due to a planet and the bisector variation[24].

---

6      http://geco.oeaw.ac.at/software.html

We performed a simulation to determine whether starspots could cause the observed shifts and shape changes of the bisectors. For DMPP-2, the macroturbulent broadening of $v_{mac}$= 8.6 km s$^{-1}$ dominates the line profile width, being greater than the estimated rotational broadening of $v$sin$i$ = 5.1 km s$^{-1}$. We used the *Doppler Tomography of Stars* imaging code, DoTS[31], to simulate line profiles using the parameters in Table 1 for a star with spots 1500 K cooler than the photosphere, with axial inclination, $i = 90°$, and an equatorial spot. We find little change in the line bisector shapes. This is expected because the Doppler information from stellar surface inhomogeneities is poorly resolved when the macroturbulent velocity is larger than $v$sin$i$. To produce parallel shifts of amplitude ~150 m s$^{-1}$, an unfeasibly large spot of radius 10° is needed. This, coupled with the low activity indicated by log($R'_{HK}$) of DMPP-2, implies the observed parallel shifts and significant changes in bisector shapes cannot be reasonably explained by cool starspots.

The residual time series of the mean deconvolved line profile, derived from 6900 absorption lines in each spectrum[32], is shown in Supplementary Fig 7. The line shifts predicted by our Keplerian solution were subtracted prior to making the time series. The BIS and the FWHM each parameterise the line profile with a single number. The trailed spectrogram presented in Supplementary Fig 7 gives information on the full shape of the line profile and its evolution in time. This can be a sensitive way to reveal patterns and may aid in placing our data in the context of future observations of this system.

## Analysis of Archival Photometry on DMPP-2

The SuperWASP archive contains observations of DMPP-2 (1SWASP J014937.88-342732.8) spanning 8 years (2006-2014), with over 60,000 individual photometric measurements. We performed searches for periodic modulations, initially using a version of the 1-dimensional CLEAN algorithm[33] which is suited to such unevenly sampled data. Potentially significant periods found from the power spectrum were then analysed using a method combining phase dispersion minimization and epoch folding[34] to precisely identify any modulation periods. This revealed power only at periods around and commensurable with 1 d, attributable to sampling artefacts. The strongest of these signals is at 1.03537 d and has an amplitude of 0.6% (Supplementary Fig 8).

The photometry was taken between the MIKE and DMPP observations, i.e., with no accumulated phase offsets due to period uncertainties. Accordingly, we folded the SuperWASP photometry on each of the periods in Table 2. None of these showed any sign of a transit at inferior conjunction of the planet, with a limit on the transit depth of $\delta < 0.6\%$. These data appear dominated by correlated noise **on a ~1 d timescale** with artefacts of approximately this amplitude throughout the folded lightcurves.

**Supplementary Table 1**: DMPP-2 stellar parameters. Atmospheric parameters from our spectral and SED analysis are included in the top section, while our isochrone fitting results are listed below.

| Parameter | Value | Reference |
|---|---|---|
| Spectral Type | F5V | [1] |
| Parallax [mas] | $7.45 \pm 0.24$ | [2] |
| Distance [pc] | $134 \pm 4$ | [2] |
| $V$ [apparent mag.] | 8.6 | [1] |
| $B-V$ [apparent mag.] | $0.463 \pm 0.015$ | [1] |
| $E(B-V)$ [mag.] | $0.026 \pm 0.008$ | This work |
| $\log(R'_{HK})$ | -5.26 (-5.36 to -5.17) | [3] |
| $T_{\text{eff}}$ [K] | $6500 \pm 100$ | This work |
| $\log g$ [cm s$^{-2}$] | $3.8 \pm 0.2$ | This work |
| $v\sin i$ [km s$^{-1}$] | $5.1 \pm 1.0$ | This work |
| $v_{\text{mac}}$ [km s$^{-1}$] | $8.6 \pm 1.0$ | This work |
| $v_{\text{mic}}$ [km s$^{-1}$] | $1.6 \pm 0.1$ | This work |
| $R_*$ [R$_\odot$] | $1.87 \pm 0.09$ | This work |
| $L_*$ [L$_\odot$] | $1.41 \pm 0.16$ | This work |
| $M_*$ [M$_\odot$] | $1.44 \pm 0.03$ | This work |
| $\log g$ [cm s$^{-2}$] | $4.1 \pm 0.04$ | This work |
| $R_*$ [R$_\odot$] | $1.78 \pm 0.09$ | This work |
| $L_*$ [L$_\odot$] | $1.27 \pm 0.16$ | This work |
| Age [Gyr] | $2.0 \pm 0.3$ | This work |

**Supplementary Table 2:** LTE atmospheric abundances of DMPP-2 with the error estimates based on the internal scatter from the number of measured lines, $n$. An 'S' in column 3 indicates that the abundance was derived with spectral synthesis. For comparison purpose, column 4 lists the solar abundances[18].

| Ion | DMPP-2 (HD 11231) log $(N/N_{tot})$ | n | Sun Log$(N/N_{tot})$ |
|---|---|---|---|
| Li I | -9.60 ± 0.05 | S | -10.99 |
| C I | -3.63 ± 0.10 | 9 | -3.61 |
| Na I | -5.66 ± 0.01 | 2 | -5.80 |
| Mg I | -4.42 ± 0.16 | 2 | -4.44 |
| Al I | -5.51 | 1 | -5.59 |
| Si I | -4.52 ± 0.17 | 26 | -4.53 |
| Si II | -4.46 | 1 | -4.53 |
| S I | -4.90 ± 0.03 | 2 | -4.92 |
| Ca I | -5.45 ± 0.07 | 17 | -5.70 |
| Sc II | -8.95 ± 0.08 | 5 | -8.89 |
| Ti I | -7.00 ± 0.08 | 25 | -7.09 |
| Ti II | -7.10 ± 0.21 | 14 | -7.09 |
| V I | -8.05 ± 0.02 | 5 | -8.11 |
| Cr I | -6.22 ± 0.08 | 28 | -6.40 |
| Cr II | -6.06 ± 0.09 | 4 | -6.40 |
| Mn I | -6.56 ± 0.07 | 9 | -6.61 |
| Fe I | -4.38 ± 0.09 | 270 | -4.54 |
| Fe II | -4.40 ± 0.08 | 44 | -4.54 |
| Co I | -7.02 ± 0.08 | 7 | -7.05 |
| Ni I | -5.70 ± 0.07 | 66 | -5.82 |
| Cu I | -7.79 ± 0.06 | 2 | -7.85 |
| Zn I | -7.51 ± 0.01 | 2 | -7.48 |
| Sr I | -8.31 | 1 | -9.17 |
| Y II | -9.78 ± 0.03 | 5 | -9.83 |
| Zr II | -9.28 | 1 | -9.46 |
| Ba II | -9.52 ± 0.06 | 3 | -9.86 |
| La II | -10.83 | 1 | -10.94 |
| Ce II | -10.38 | 1 | -10.46 |
| Nd II | -10.62 | 1 | -10.62 |
| Sm II | -11.03 | 1 | -11.08 |
| Eu II | -11.33 | 1 | -11.52 |

**Supplementary Table 3:** Maximum likelihood statistics, false alarm probabilities and *a posteriori* parameters for three different models for DMPP-2 RVs with FWHM and BIS correlation terms. See Table 1 of main text for further details. The same periods as in Table 1 (main text) are still detected with low FAP.

| Parameter | P95-97 + P98 ($\gamma_{P95+P97}, \gamma_{P98}$) **FWHM correlation** | P95-P98 ($\gamma_{HARPS} = \gamma_{P95+P97+P98}$) **BIS correlation** | P95-97 + P98 ($\gamma_{P95+P97}, \gamma_{P98}$) **BIS correlation** |
|---|---|---|---|
| **Maximum likelihood parameters** | | | |
| FAP | $7.2 \times 10^{-8}$ | $1.2 \times 10^{-6}$ | $2.9 \times 10^{-7}$ |
| $\Delta \log L$ | 30.6 | 25.2 | 28.9 |
| *Maximum a posteriori parameters* | | | |
| $P$ [d] | 5.2065(5.2040 - 5.2084) | 5.9978(5.9946 – 5.9999) | 5.2048 (5.2029 - 5.2067) |
| $K$ [ms$^{-1}$] | 43.92(38.53 - 47.46) | 38.81(354.98 - 44.56) | 40.55 (35.48 - 44.54) |
| $e$ | 0.087(< 0.093) | 0.031(< 0.068) | 0.065 (< 0.081) |
| $\lambda = M_0 + \omega$ [deg] | 13.71(6.97 - 20.19) | 32.08(21.12 - 38.33) | 12.76 (6.57 – 21.58) |
| $\gamma_{HARPS}$ [ms$^{-1}$] | | -6.88(-9.32 - -3.45) | |
| $\gamma_{P95+P97}$ [ms$^{-1}$] | -17.14(-19.36 - -13.06) | | -15.66 (-17.49 - -11.30) |
| $\gamma_{P98}$ [ms$^{-1}$] | 5.74(-0.22 - 10.58) | | 5.70 (2.07 - 11.87) |
| $\sigma_{HARPS}$ [ms$^{-1}$] | | 17.46(17.02 – 21.23) | |
| $\sigma_{P95+P97}$ [ms$^{-1}$] | 12.03(11.15 - 16.53) | | 13.00 (12.74 – 17.32) |
| $\sigma_{P98}$ [ms$^{-1}$] | 17.12(16.55 – 23.99) | | 17.12 (16.93 – 24.79) |

**Supplementary Table 4:** Summary of the December 2018 status of 39 DMPP targets[35] with log $R'_{HK}|_{max}$ < -5.1.

| Target(s) | Nobs | Status |
|---|---|---|
| DMPP-1 | 148 | Compact multi-planet system |
| DMPP-2 | 56 | Planet orbiting pulsating star |
| DMPP-3 | 93 + 8 archival | Circumprimary super-Earth in binary star system. |
| P1-N *(DMPP-4)* | 153 | Preliminary: At least two signals, two or three low mass planets with $P_{orb}$ < 3d. GIARPS observations pending. Under analysis. |
| P1-S *(DMPP-5)* | 81 | Preliminary: 2 super-Earths $P_{orb}$ = 3.6d, 6.4d; $M_P \sin i$ = 3.5 $M_\oplus$, 3.8 $M_\oplus$ (no correlations w/ activity indicators). Possible ~90 d low mass giant. Under analysis. |
| LP-S | 10 +82 archival | Long period 4 $M_J$ giant planet, more DMPP observations needed to search for short period planets. |
| P2-S, P3-S | 52, 81 | Evidence for low amplitude < 2.5 d period Keplerian signals with FAP < 1% |
| 2 targets | < 60 | ~2 $M_\oplus$ planets in sub-day orbits appear to be excluded |
| 5 targets | < 60 | Unclear RV behaviour. More observations required. |
| 1 target | < 60 | Unclear RV behaviour. Probable pulsator, with pulsation-driven RV variability. |
| 22 targets | 0 | Observations required. |
| 1 target | 7 | log $R'_{HK}$> -5.1 from our spectra. Dropped. |

## Supplementary Information references

## Supplementary Figures

**Supplementary Figure 1:** HARPS-TERRA RVs for four FGK RV standard stars. Points with colour-coded 1-σ error bars, as a function of signal-to-noise in HARPS order 60 ($SNR_{60}$), are shown. The best fit (see Equation 2) is shown with the black line (dashed lines represent the fit uncertainties). The correction derived for M dwarfs by Berdinas et al.[36] is also indicated with the red line (see Equation 1). The data for each star are offset to RV = 0.0 at $SN_{60} = 115$.

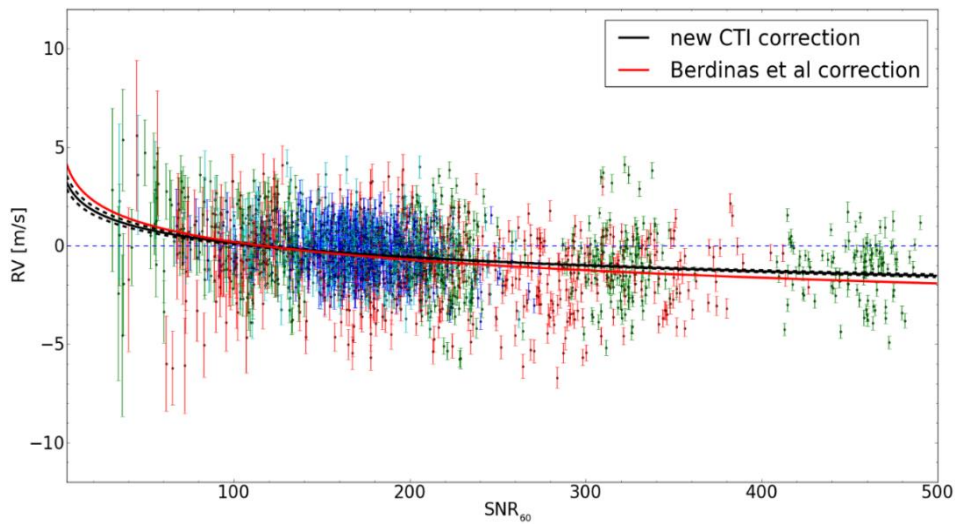

**Supplementary Figure 2:** Calibration of the CTI effect for SOPHIE (as in Supplementary Fig 1), showing measured RV a function of SNR. Top panel: Best fit comparing our new calibration (see Equation 3) with the earlier Santerne et al. SOPHIE calibration[37]. Lower two panels: residuals for our new calibration and the previous SOPHIE calibration.

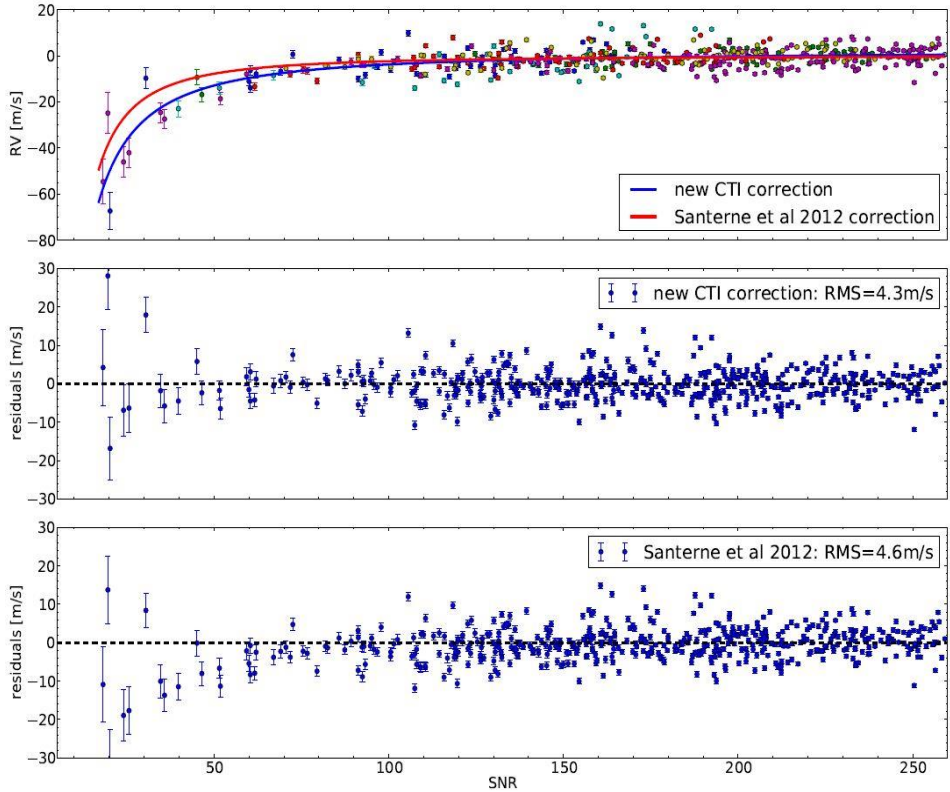

**Supplementary Figure 3**: Pseudo spectral energy distributions (pSED) from SOPHIE observations of a K star target at a variety of airmasses. Differential atmospheric refraction losses are apparent from changes in slope. Upper panel: Summed counts in each echelle order are normalised to the reference HARPS order 59, indicated by "R". Lower panel: As above, but divided by the first observation (blue lines) and labelled "T". A linear fit within the blue highlighted region (HARPS orders 54 to 64) is used to determine κ.

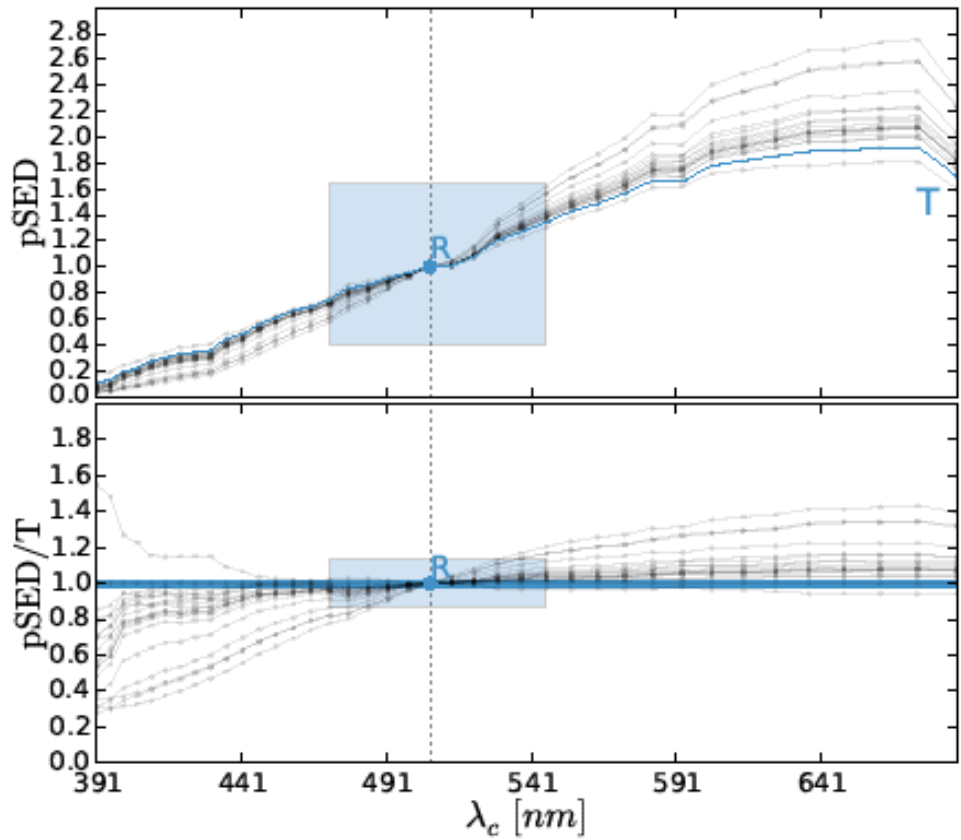

**Supplementary Figure 4:** Periodograms of activity indices for DMPP-2. (a) The window function and likelihood periodograms for the (b) BIS, (c) FWHM and (d) S-index of the HARPS data. The positions of the RV signal peaks at $P = 5.998$ d and $5.206$ d are highlighted (vertical dashed lines), and the 10, 1 and 0.1 FAP thresholds are shown (horizontal dashed lines). Significant peaks are present in the BIS and FWHM, but the clusters of peaks are distinct from the RV periodicities. There is no significant variability in the S-index.

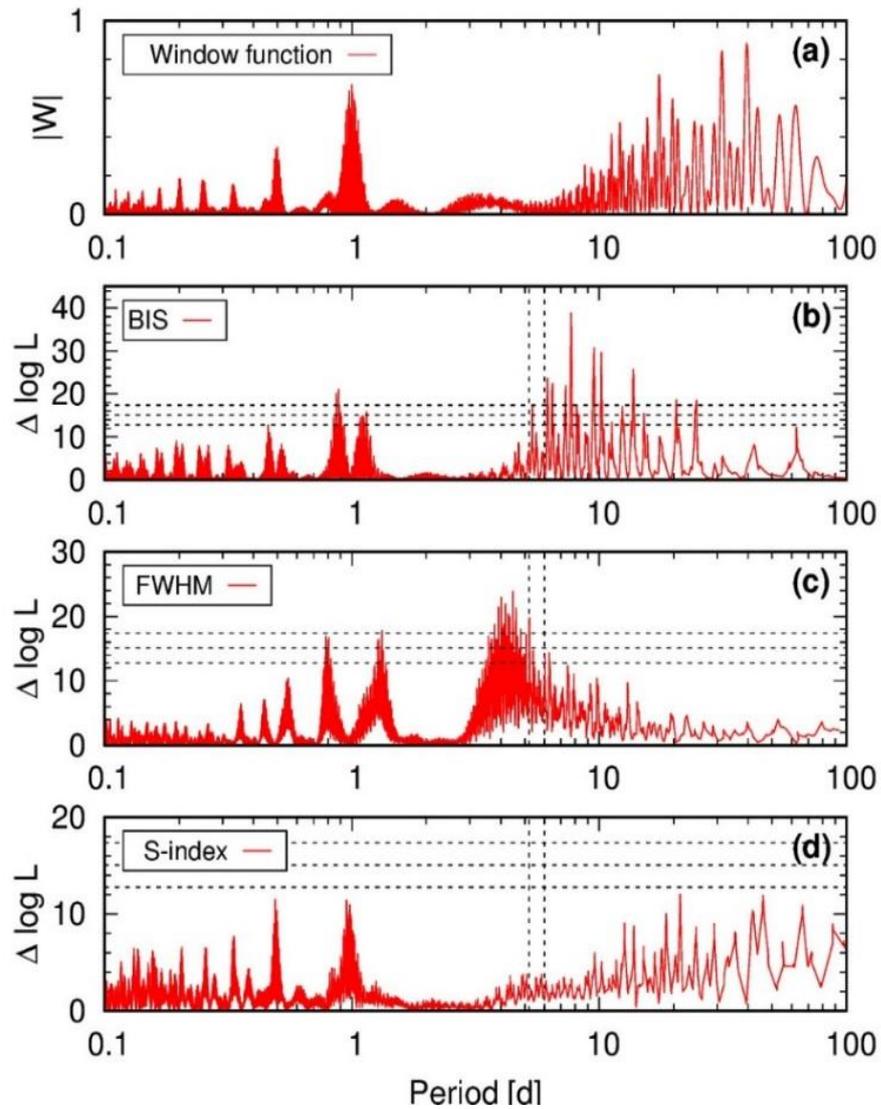

**Supplementary Figure 5:** DMPP-2 HARPS activity-RV correlations. Plots are shown for (a) BIS vs RV (b) BIS vs residual RV and (c) BIS vs RV and (d) S-index vs RV. For each parameter the mean has been subtracted. Estimated RV and activity parameter 1-σ uncertainties are shown.

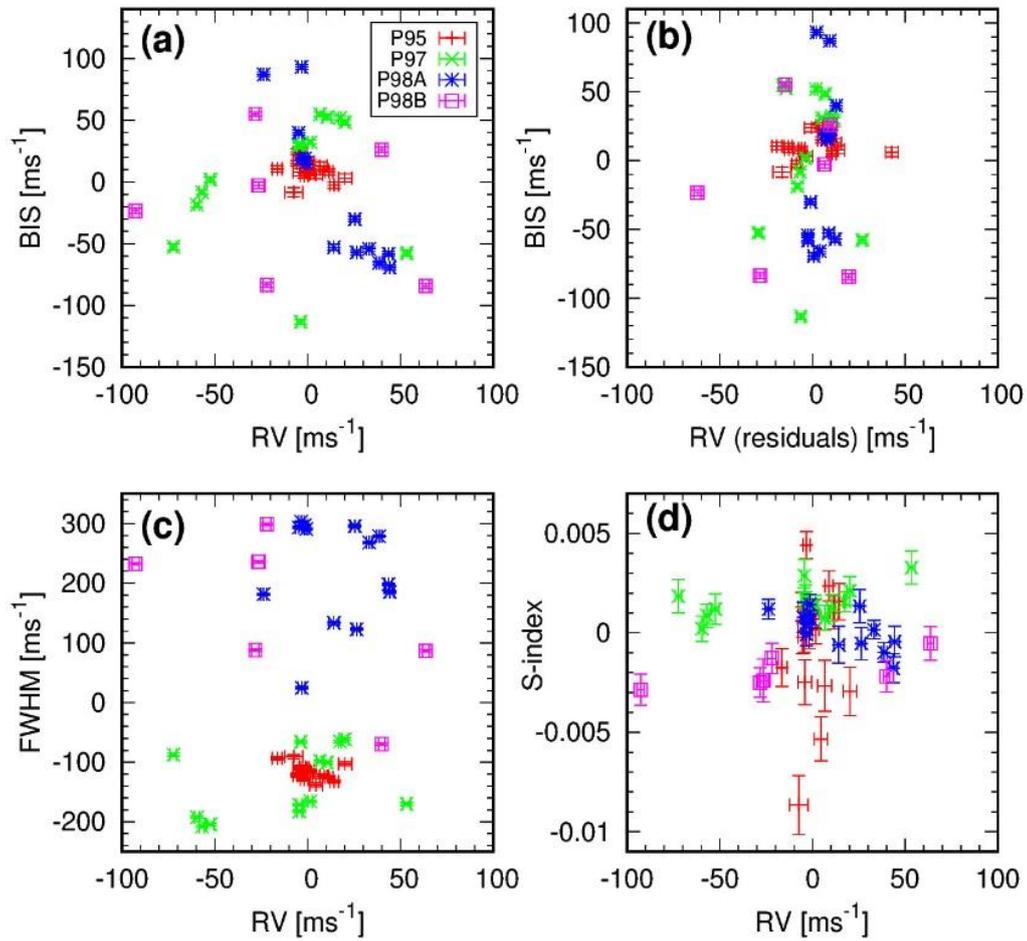

**Supplementary Figure 6:** Line bisectors of DMPP-2 for each HARPS epoch. (a) The bisectors are colour-coded by phase of the best fit $P_{orb}$ = 5.205 d orbital solution that fits for $\gamma_{P95+P97}$ and $\gamma_{P98}$ (column 2 of Table 2). Quadrature points of the orbit are red (phase 0.25) and blue (phase 0.75), while conjunctions are black. Note that using the 5.998 d solution (fitting with a single $\gamma_{HARPS}$) gives very similar results. (b) Bisectors shifted by the measured TERRA velocity at each epoch.

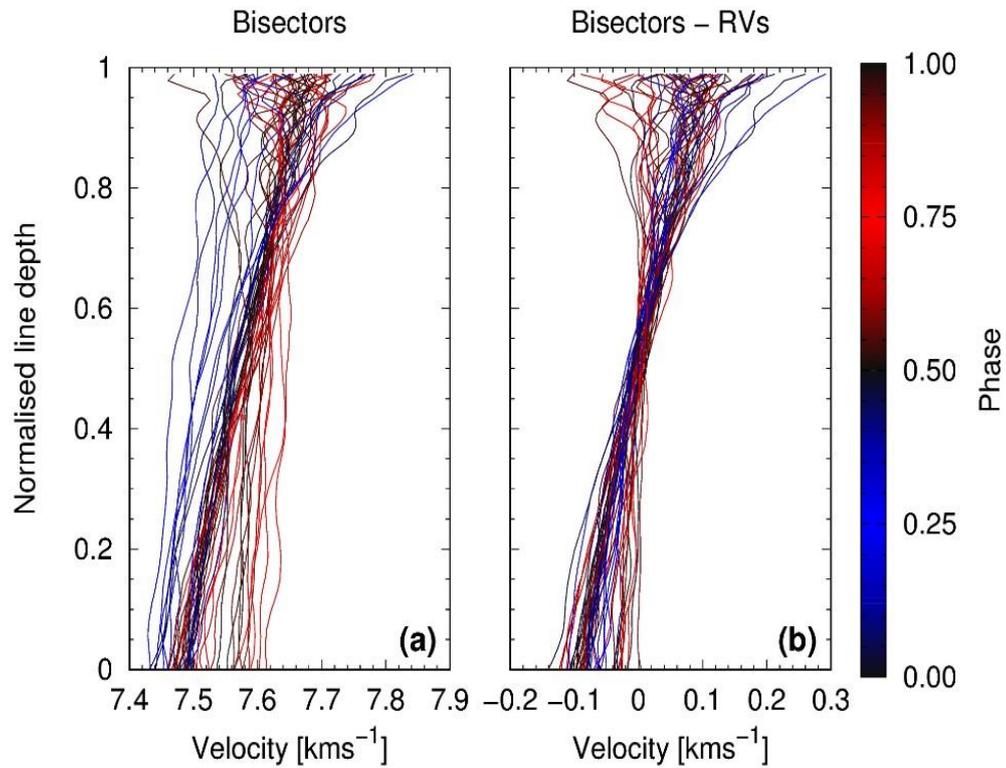

**Supplementary Figure 7:** Trailed residual spectrum of least squares deconvolved absorption line profiles for our 49 HARPS spectra. The residuals (in normalised continuum units) are calculated by subtracting the mean profile from each profile in turn. Observations on a given night are separated with thin dashed lines. Observing runs P98A and P98B are separated by a thick line. The vertical dashed lines represent the profile width of the combined broadening mechanisms including *v*sini (see Supplementary Table 1).

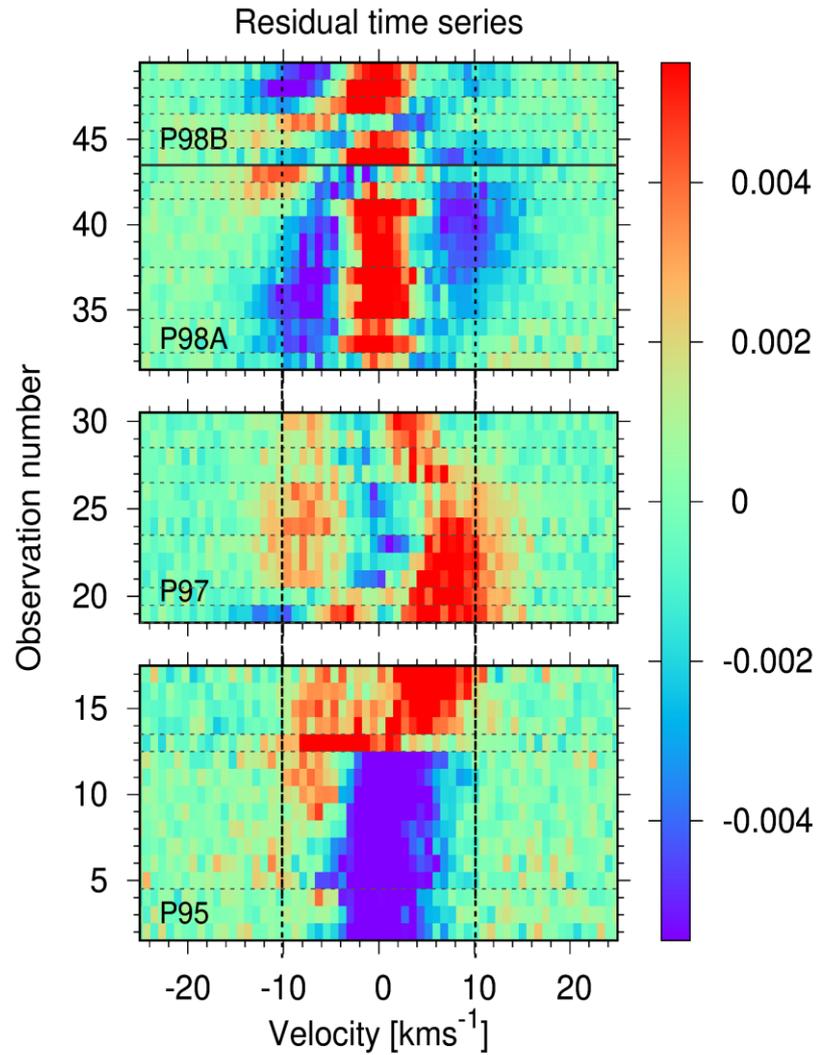

**Supplementary Figure 8:** SuperWASP observations of DMPP-2 folded on the strongest period found, 1.03537d. Individual observations are plotted as points without error bars. The red points show the phase binned data and associated propagated errors. The bin width is 0.01 in phase, with ~400 points per bin. The detected modulation could be due to sampling artefacts; the 0.6% amplitude is well within the scatter.

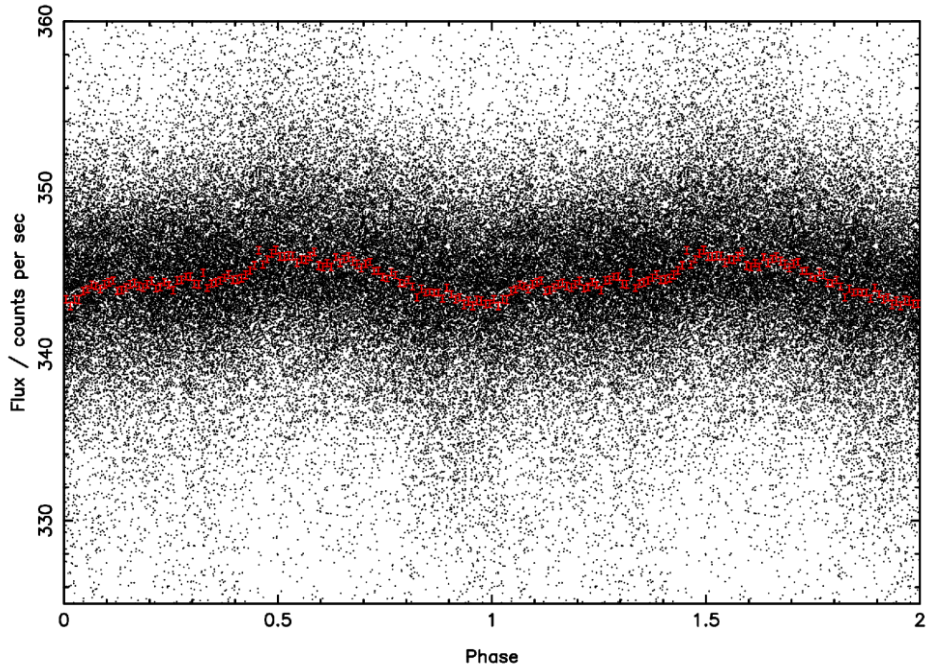

**Supplementary Figure 9:** An example sensitivity calculation from Monte Carlo simulations incorporating 94 existing **HARPS observations** and **10** proposed **ESPRESSO** observations of a V ≈ 8 target.

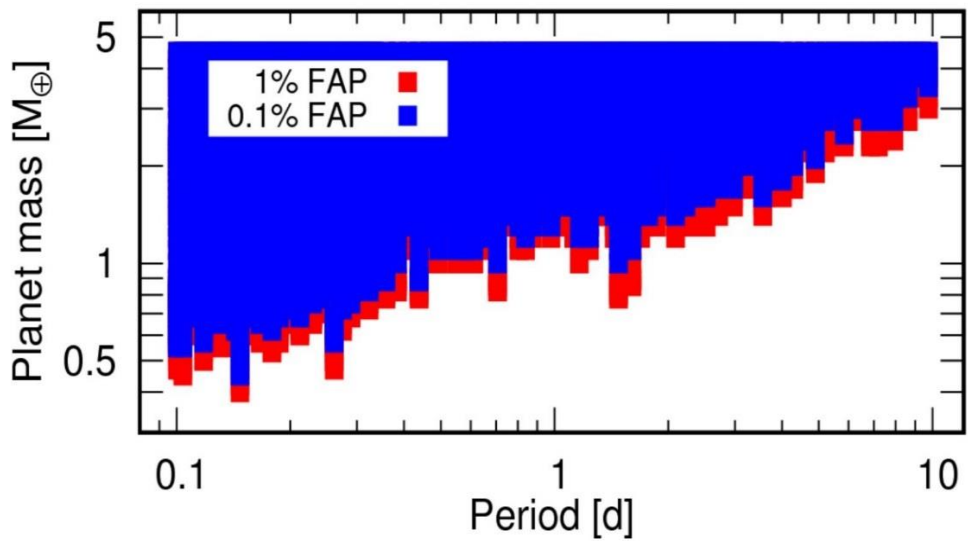